\documentclass[aps,prb,twocolumn,superscriptaddress,showkeys]{revtex4-2} %Physical Review B

\usepackage{array,multirow}
\usepackage{graphicx}
\usepackage{xcolor}
\usepackage{ulem}

\makeatletter

\newcommand{\pcom}[1]{#1}
\newcommand{\gcom}[1]{#1}
\newcommand{\dcom}[1]{}

\begin{document}

%\email for Physical Review B \ead for 2D Materials
%\affiliation for Physical Review B \address for 2D Materials

%Title for 2D Materials
%\title[Thickness and electric field effects in TMDs]{Bands tuning in transition metal dichalcogenide heterostructures: the interplay between thickness and electric field}
%Title for Physical Review B
\title{\pcom{Tunable electronic and magnetic properties of thin Nb$_3$I$_8$ nanofilms: interplay between strain and thickness}}
%\title{Strain-tunable electronic and magnetic properties of Nb$_3$I$_8$ multilayer}
\author{Giovanni Cantele}
\email{giovanni.cantele@spin.cnr.it}
\email{giovanni.cantele@unina.it}
\affiliation{CNR-SPIN, c/o Complesso Universitario M. S. Angelo, via Cintia 21, 80126, Napoli, Italy}
\author{Felice Conte}
\affiliation{Dipartimento di Fisica ``E. Pancini'', Universit\`a degli Studi di Napoli ``Federico II'', Complesso Universitario M. S. Angelo, via Cintia 21, 80126, Napoli, Italy}
\affiliation{CNR-SPIN, c/o Complesso Universitario M. S. Angelo, via Cintia 21, 80126, Napoli, Italy}
\author{Ludovica Zullo}
\altaffiliation{Present address: Department of Physics, University of Trento, Via Sommarive 14, 38123 Povo, Italy}
\affiliation{Dipartimento di Fisica ``E. Pancini'', Universit\`a degli Studi di Napoli ``Federico II'', Complesso Universitario M. S. Angelo, via Cintia 21, 80126, Napoli, Italy}
\author{Domenico Ninno}
\affiliation{Dipartimento di Fisica ``E. Pancini'', Universit\`a degli Studi di Napoli ``Federico II'', Complesso Universitario M. S. Angelo, via Cintia 21, 80126, Napoli, Italy}
\affiliation{CNR-SPIN, c/o Complesso Universitario M. S. Angelo, via Cintia 21, 80126, Napoli, Italy}
\date{\today}

\begin{abstract}
The study of novel 2D platforms implementing magnetism in tunable
van der Waals (vdW) homo- and hetero-structures paves the way to innovative
spintronics and magnetic devices. In this study, we unravel the intriguing
properties of few-layer Nb$_3$I$_8$ vdW \pcom{nanofilms} from first principles, showing
\pcom{how and to what extent} 
specific magnetic orderings can be tuned using several
degrees of freedom, such as film thickness, stacking geometry, 
and strain or even a combination of them. All these aspects are explored here, giving a 
comprehensive view of this novel and promising magnetic material.
\end{abstract}

%keywords for Physical Review B
%
%keywords for 2D Materials
%\vspace{2pc}
%\noindent{\it Keywords}: van der Waals heterostructures, transition metal dichalcogenides, electric field, nearly free electron states, tuning of transport properties \\
%\\

% Uncomment if a separate title page is required
\maketitle

%Section for 2D Material
% Uncomment for Submitted to journal title message
%\submitto{\TDM}
% 
% For two-column output uncomment the next line and choose [10pt] rather than [12pt] in the \documentclass declaration
%\ioptwocol
%

\section{Introduction}
\label{sec:intro}
In the last years, two-dimensional (2D) materials have been attracting tremendous research interest.
Their charming properties, induced by the combination of 
surface effects and quantum confinement, can be exploited to realize 2D platforms-based electronic
devices for a wide range of technological applications 
\cite{Liu2021, Akinwande2019, Liu2020}.
The out-of-plane van der Waals (vdW) interactions allow the integration of deeply different 2D materials, 
that can thus be viewed as \pcom{elementary} building blocks of new \pcom{heterostructures}
showing novel properties and exotic phenomena unavailable in the single-layer constituents.
An unprecedented number of degrees of freedom, such as order and number of layers \cite{C5NR03895B, santos, PhysRevMaterials.1.014002}, the twist angle among them \cite{Cao2018, A1, 195419,TBG_PRR}, and the distance between two consecutive single layers (interlayer distance) \cite{C5RA20882C, C5NR06293D} can be tuned
to achieve this goal.

\pcom{Among these degrees of freedom, strain deserves a special mention, since it offers a viable approach
to engineer the electronic, magnetic, and optical 
properties \cite{Cao2020, Sando2013, Fei2014, Castellanos-Gomez2013} of 2D nanostructures.
The reduced atomic coordination and the enhanced flexibility and elasticity that
2D materials often exhibit when compared with most three-dimensional (3D)
crystals \cite{Lee385, Bertolazzi2011} have boosted the development of this new field,
referred to as ``straintronics''.}

\pcom{As far as ground-state magnetism is concerned, several avenues have been devised so far
to gain a precise and predictable control over the magnetic states in a number of recently discovered 2D materials.}
%\dcom{ Electrostatic doping applied to bilayer CrI$_3$ significantly affects the interlayer exchange coupling, providing a phase transition from the layered AFM state of two FM layers with opposite spin directions to an FM state \cite{Jiang2018, Huang2018}. An out-of-plane pressure of 2.70 GPa on bilayer CrI$_3$ destroys the AFM state to favor an FM ordering \cite{Li2019, Song2019}, whereas an out-of-plane pressure of 2.45 GPa on trilayer CrI$_3$ can create coexisting domains of three phases, one ferromagnetic and two antiferromagnetic, or can induce an FM ordering \cite{Song2019}. Strain engineering was shown to affect the magnetic properties of several 2D magnets, such as FM-AFM phase transition of monolayer CrI$_3$ \cite{C8CP07067A} and transition metal trichalcogenides \cite{PhysRevB.101.085415}, valley polarization and metal-semiconductor and/or magnetic phase transitions of CoSe structures \cite{PhysRevB.102.224422}, the superexchange interactions and the Curie temperature of Cr$_2$Ge$_2$Te$_6$ \cite{PhysRevApplied.12.014020}.}
\pcom{For example, it has been argued that charge doping and/or external strain can significantly
affect and modify
magnetic phase transitions and exchange interaction~\cite{Memarzadeh_2021}, and are being
explored as feasible routes to tune the magnetic ordering in both monolayer ad multilayers systems~\cite{Jiang2018,Huang2018,Li2019,Song2019,C8CP07067A,PhysRevB.101.085415,PhysRevB.102.224422,PhysRevApplied.12.014020}.}

Since the recent discovery of intrinsic magnetism in monolayers of CrI$_3$ \cite{Huang2017},
the exploration of 2D intrinsic magnetic materials has 
been exponentially increasing \cite{doi:10.1063/5.0025658}. \pcom{Among the many, we can mention
CrI$_3$ and Cr$_2$Ge$_2$Te$_6$ \cite{Gong2017}, or those exhibiting high 
Curie temperature $T_C$, such as V$_3$I$_8$ 
\cite{C9CP00850K}, MnS$_2$ \cite{C3CP55146F}, VSe$_2$ \cite{Fuh2016}, and Nb$_3$I$_8$ \cite{Nb3I8_PRR}, all inheriting the magnetic ordering arising from 
the transition metal $d$-orbitals.}
In particular, 2D platforms combining ferromagnetism with room $T_C$ 
and conventional semiconductors open \pcom{new avenues} to \pcom{implement} spintronic
applications based on the use of both the charge and the spin degrees of freedom\pcom{, for example, in
next-generation} quantum logic chips and nonvolatile magnetic memories with increased densities \cite{RevModPhys.76.323, Wolf1488}.

Nb$_3$I$_8$ is a recently synthesized 2D material \cite{Nb3I8_synthesis} with a predicted ferromagnetism at the room temperature ($T_C\sim307$ K) and a layer-dependent magnetism, being ferromagnetic (FM) in monolayer form and
antiferromagnetic (AFM) in bilayer and trilayer forms \cite{Nb3I8_PRR}. Also, Nb$_3$I$_8$ monolayer actually is
a ``ferrovalley'' material, because it exhibits an intrinsic spontaneous valley polarization of 107 meV and, thus, 
the anomalous valley Hall effect without external tunings \cite{Nb3I8_valley}. All these observations make
Nb$_3$I$_8$ an ideal candidate for spintronics and valleytronics applications, \pcom{as even revealed by
recent experimental and theoretical studies \cite{arxiv.2203.10547}}.
%All these evidences clearly bring out that strain engineering reveals as one of the most promising ``control
%knobs'' to engineer the electronic and, to an even larger extent, magnetic properties of 2D magnetic materials.
\pcom{Nevertheless, a comprehensive explanation of the intriguing electronic and magnetic properties of
few-layer Nb$_3$I$_8$ is still missing, so the present work aims to unreveal its potentialities for
spintronics applications contributing to the search of novel materials with
improved or new functionalities.}
%\dcom{With this motivation, the} The present
%work is focused on a detailed analysis of the intriguing electronic and magnetic properties of the less explored Nb$_3$I$_8$, to unravel its
%potentialities for spintronic applications \pcom{and in the search of novel materials with
%improved or new functionalities}.
State-of-the-art first-principles calculations in the framework of density functional theory
(DFT) are carried out to bring out the effects of the (in-plane) biaxial strain on the electronic and magnetic
properties of Nb$_3$I$_8$ in monolayer, bilayer, and trilayer forms. All these systems are investigated in two different
stacking geometries, to establish a possible interplay between thickness, stacking, strain, and magnetism.
\pcom{In particular, we will assess and describe the competition and relative stability between 
different magnetic phases and the band-gap dependence on the applied strain.}
%\dcom{-induced band gap engineering, showing intriguing properties, such as possible transitions between magnetic
%states. As it will be better detailed in the next, due}

\pcom{Due} to the peculiar nature of Nb$_3$I$_8$, accurate
first-principles calculations should definitely take into account two fundamental interactions, that is, vdW interlayer interaction
(by an appropriate choice of the exchange-correlation functional) and on-site Coulomb repulsion for  Nb 4$d$-electrons (for example, using a DFT+$U$
scheme). Both are carefully included in our analysis, accompanied by a systematic study of the convergence of the presented results with
respect to all parameters involved in the calculations. \pcom{This will be better detailed in the next section}.

The paper is organized as follows. In Sec. \ref{sec:methods} we present the computational methods and technical details of our calculations. In Sec. \ref{sec:results} we show and discuss the electronic and magnetic properties of Nb$_3$I$_8$ films, with special focus on their dependence on
the thickness, strain and stacking geometry. Finally, in Sec. \ref{sec:IV}, we summarize our results and draw some conclusions.

\begin{figure*}
(a)\includegraphics[scale=0.12]{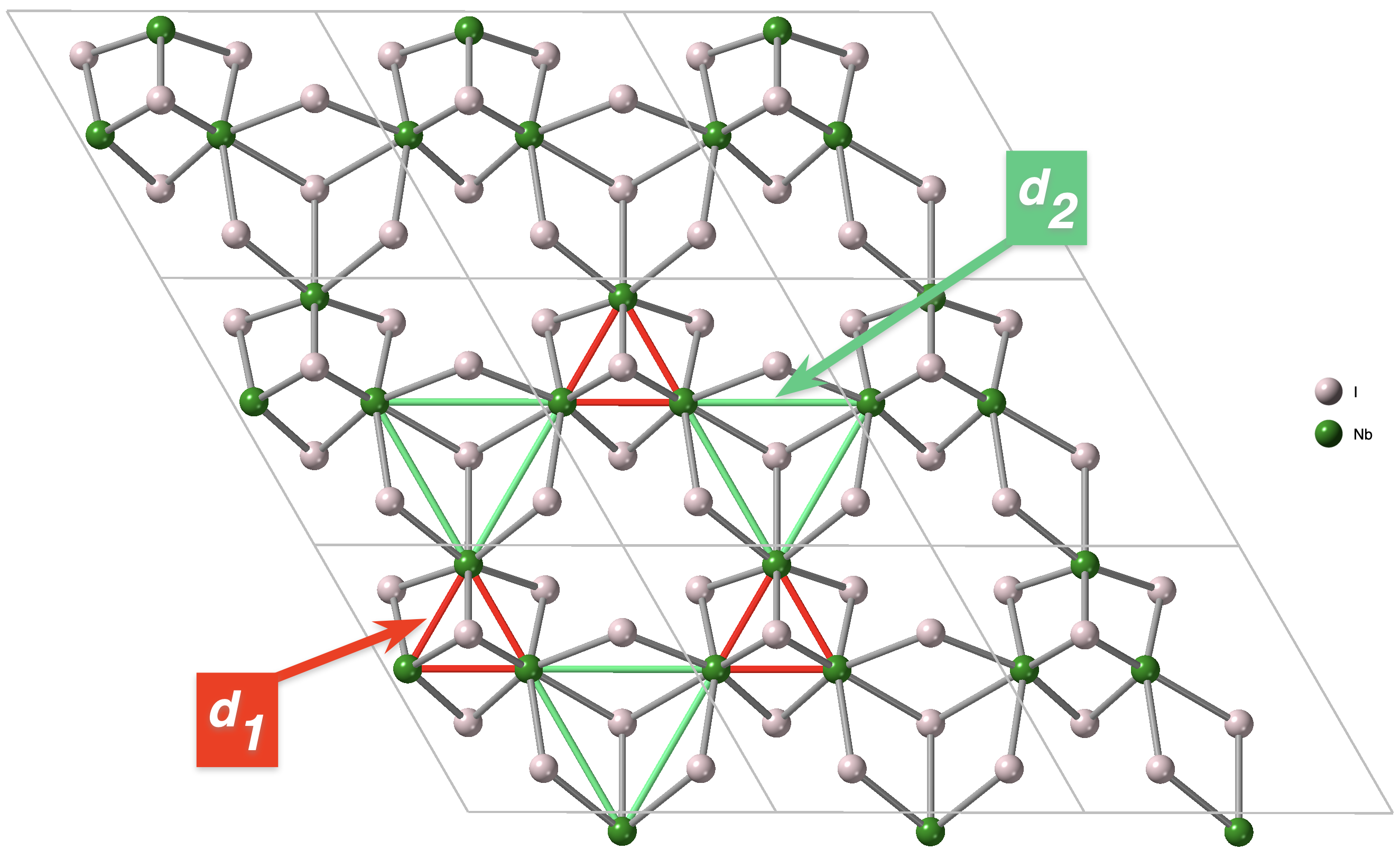} \hfill (c)\includegraphics[scale=0.253]{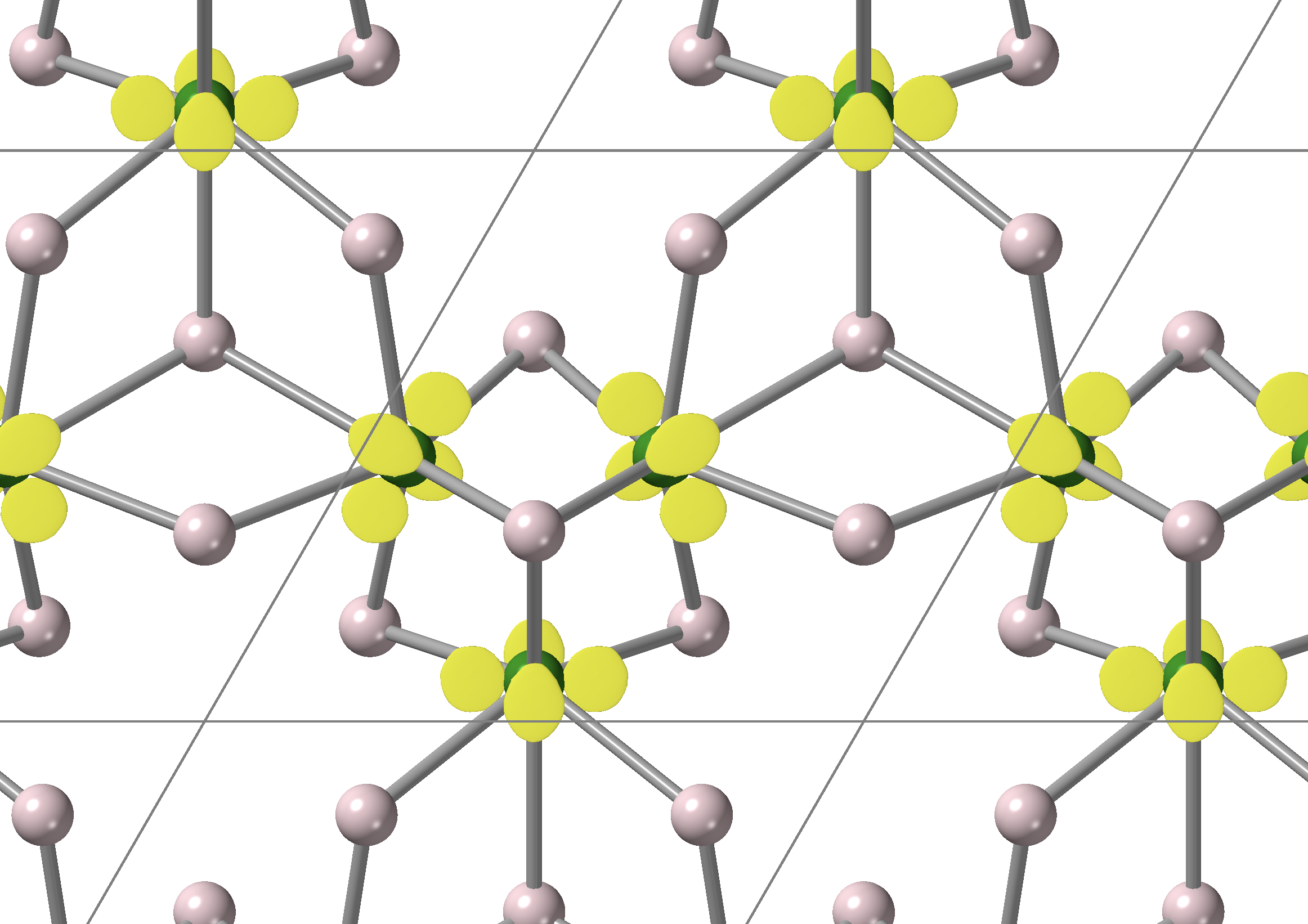}\vspace*{0.8cm}

(b)\includegraphics[scale=0.1]{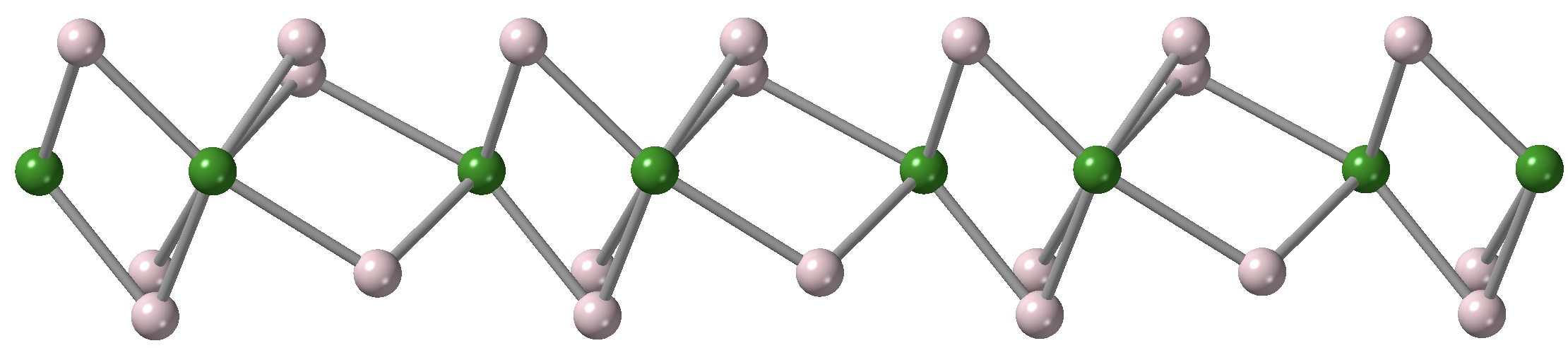}\hspace*{10cm}
    \caption{Top (a) and side (b) view of a Nb$_3$I$_8$-1L $3\times3$ supercell. Green and pink spheres represent
Nb and I atoms, respectively. The irregular Kagom\'e lattice \pcom{formed by} Nb atoms can be recognized in the \pcom{top view,
together} with the first-neighbor $d_1\simeq2.95$ {\AA} and second-neighbor $d_2\simeq4.65$ {\AA} Nb-Nb distances \pcom{evidenced
by red and green bonds respectively. This gives rise to alternating, inequivalent triangles \cite{arxiv.2203.10547,Sun2022}.
\pcom{The ground-state spin polarization distribution $\rho_\uparrow-\rho_\downarrow$ is shown in (c) with the yellow isosurfaces, corresponding 25\% of the maximum value.}}}
    \label{fig:structure}
\end{figure*}

\begin{figure}
\includegraphics[scale=0.28]{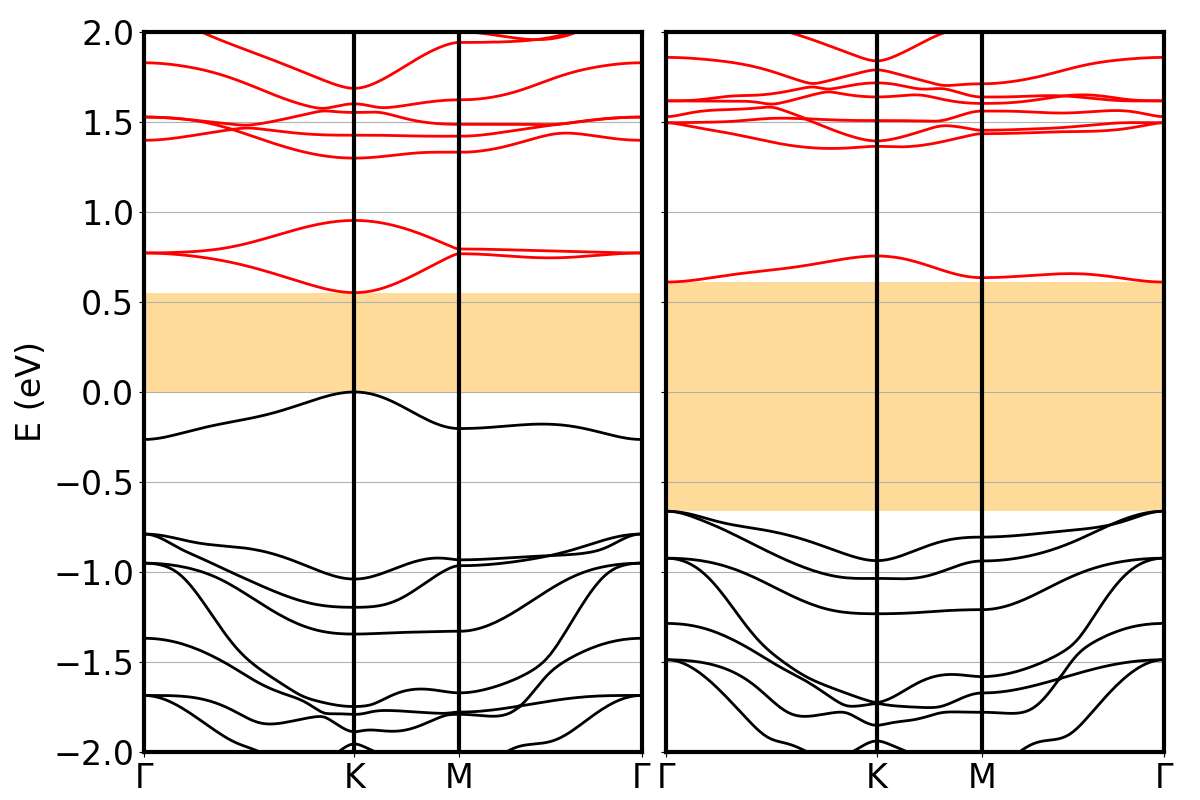}
    \caption{Spin-up (left) and spin-down (right) band structure along the $\Gamma-K-M-\Gamma$ path in the first BZ 
    of Nb$_3$I$_8$-1L. Zero energy corresponds to the top valence band. The band structure shows a semiconducting
    behavior, with the spin-up and spin-down energy gaps highlighted by a shaded orange region \pcom{and occupied
    (unoccupied) bands depicted in black (red).}}
    \label{fig:1Lunbands}
\end{figure}

\section{Methods}
\label{sec:methods}
All calculations are performed using DFT as implemented in the \textsf{Quantum-ESPRESSO} package (version 6.6) \cite{Giannozzi_2009, Giannozzi_2017, Giannozzi_2020}. The generalized gradient approximation (GGA) is used with projector-augmented wave (PAW) 
pseudopotentials \cite{DALCORSO2014337} based on the Perdew-Burke-Ernzerhof (PBE) exchange-correlation functional \cite{PhysRevLett.77.3865} to represent the 
atomic cores \cite{PSEUDO}. The plane wave basis set is truncated using a cutoff energy of 60 Ry for the plane waves and 480 Ry to represent the charge density in all calculations. An adequate vacuum space of $\simeq 20$ {\AA} \pcom{is} set between periodic replicas along the direction orthogonal to the planes (assumed to be the 
$z$ direction), in order to avoid spurious interactions induced by the periodic boundary conditions. The Brillouin zone (BZ) of the $1\times1\times1$
unit cell was sampled using an 8$\times$8$\times$1 Monkhorst-Pack $k$-point grid \cite{Monkhorst}, grids for $n\times n\times1$ supercells were scaled
accordingly \pcom{to guarantee the same accuracy with respect to the $k$-point grid of all calculations}. 
These $k$-point grids have been used for both structural relaxation and total energy calculations.

The vdW interaction has been self-consistently accounted for using the 
\textsf{rev-vdW-DF2} \cite{B86R} exchange-correlation functional, that has been proven to 
be successful in the description of 2D vdW heterostructures with an hexagonal
lattice \cite{TBG_PRR} and, in particular, provides a good agreement with the 
available experimental data on Nb$_3$I$_8$ \cite{Nb3I8_PRR}.
\pcom{The supplemental material (SM), in Sec. I, reports typical intralayer binding curves calculated
within this approach, characterized by binding energies that are typical of vdW systems.}

Based on previous reports~\cite{C6NR07231C},
the on-site Coulomb repulsion of Nb 4$d$ electrons\pcom{, responsible, as we shall see, of the
magnetic behavior of the material,} is taken into account 
by means of the DFT + $U$ method \cite{LDAU1, LDAU2, LDAU3, PhysRevB.71.035105} with $U=2$ eV.
\pcom{The SM, in Sec. III, contains
a number of tests showing that, upon changing $U$ within a reasonable range, the reported properties do not
exhibit significant variations. In particular, we have verified the stability of
the ground-state magnetic ordering, that will be discussed in Sec. 
\ref{sec:results}, against $U$. Occupied Kohn-Sham levels do show changes within 0.1-0.2 eV, whereas unoccupied levels
shift by much larger energies. This allows to conclude that the main conclusions of the present paper should not
be affected by the chosen value of $U$.}

\pcom{The in-plane lattice parameter of}
all the considered systems \pcom{has been optimized using }
spin-polarized calculations. The atomic positions have been fully optimized by means of the Broyden-Fletcher-Goldfarb-Shanno (BFGS) algorithm \cite{B, F, G, S}, 
with a convergence \pcom{threshold of 10$^{-5}$ Ry on the total energy difference between consecutive
structural optimization steps and of 10$^{-4}$ Ry/Bohr on} all 
components of all the forces acting on the atoms.

In-plane, biaxial strain $\epsilon=(a-a_0)/a_0$, where $a_0$ is the \pcom{optimized in-plane} lattice parameter and
$a$ is the lattice parameter \pcom{of the strained system}, has been applied to the considered structures, ranging from $-7.5$\% to 7.5\% in steps of 2.5\%. Thickness effects
have been studied by considering monolayer, bilayer, and trilayer \pcom{nanofilms},
which will be referred to as Nb$_3$I$_8$-$n$L \pcom{with} $n$ = 1, 2, 3, 
respectively.

\dcom{Moreover, the} \pcom{The} effect of the stacking geometry has been \pcom{assessed} 
for \pcom{Nb$_3$I$_8$-$2$L and Nb$_3$I$_8$-$3$L,} by comparing 
the bulk stacking, that is the natural order of the layers as they would be arranged in the
bulk form, and the \pcom{AA} stacking, constructed by piling different layers with the same planar coordinates.
%\dcom{The latter will be referred to as AA stacking for both $n=2$ and $n=3$.}

The ground-state analysis is performed on the
non magnetic (NM) and several magnetic states for each system.
In particular, since it turns out that Nb$_3$I$_8$-1L has a magnetic ground state with
an in-plane FM ordering of the spins, \pcom{the inter-layer}
magnetic ordering in thin \pcom{nano}films with two or three layers has \pcom{to be carefully investigated}.
Indeed, while preserving the in-plane FM ordering, \pcom{the} inter-layer interaction might induce \pcom{either}
FM or AFM out-of-plane ordering of the spins, thus giving rise to \pcom{different} spin ``stacking sequences'',
such as $\uparrow\uparrow\dots$ or $\uparrow\downarrow\uparrow\dots$.
\pcom{This will be better detailed in the next.}

\pcom{Spin-orbit coupling (SOC) has not been included in the calculations, because it does not appear to
significantly affect the main conclusions of the paper. An example is shown in Sec. II of the SM,
that allows to conclude that the calculated band structures with the inclusion of the SOC do not
differ from those calculated in the absence of it.}

To conclude this section, it should be pointed out that, mostly as far as the magnetic properties are concerned,
accurate convergence tests
with respect to the calculation parameters (BZ sampling, plane wave and charge density cut-offs, and so on) are
needed. In this respect,
we refer to the appendix of Ref. \cite{Nb3I8_PRR}, where a detailed discussion of the convergence of the magnetic
stability, band structure,
Curie temperature may be found.

\section{Results and discussion}
\label{sec:results}
\subsection{Unstrained \dcom{systems} \pcom{nanofilms} }
\label{ssec:unstrained}
\subsubsection{\pcom{Bulk Nb$_3$I$_8$}}
Nb$_3$I$_8$ is a layered transition metal halide belonging to the family of Nb$_3$X$_8$ (X = Cl, Br, I) crystals,
having six layers per unit cell (u.c.) in its 
bulk structure (space group $D^5_{3d}-R\overline{3}m$, No. 166) \cite{hulliger}. Each single layer shows an I-Nb-I
sandwich structure, where the Nb atoms \pcom{are arranged into an irregular Kagom\'e lattice formed by 
triangular Nb$_3$ clusters with side $d_1\sim$2.96 {\AA}, separated by a $d_2\sim$4.65 {\AA} distance, as
sketched in Fig.  \ref{fig:structure}(a).}
\pcom{As evidenced in Fig.  \ref{fig:structure}(b), each} Nb atom is covalently bonded to a distorted octahedral environment
of I atoms, \pcom{that in turn form a top and a bottom layer that are not equivalent}.

\subsubsection{\pcom{Nb$_3$I$_8$-1L}}
Nb$_3$I$_8$-1L
is an intrinsic 2D cluster magnet \cite{Nb3I8_PRR, Nb3I8_valley}, since the magnetic moment \pcom{per unit cell}
of 1 $\mu_B$ \pcom{is distributed over} the Nb$_3$ clusters and results from the seven electrons shared by the Nb atoms.
\pcom{This is depicted in Fig. \ref{fig:structure}(c), showing the spin polarization isosurfaces in the FM ground state.}
\pcom{The NM solution is separated by an \dcom{excitation} energy of $\sim$174 meV from the ground state,}
that protects the magnetic ground state from the thermal fluctuations even at high temperatures.
\pcom{Such a circumstance is supported by the calculations reported}
\dcom{Indeed, as calculated} in Ref. \cite{Nb3I8_PRR}, \pcom{where it is shown that such}
system exhibits a Curie temperature $T_C\sim307$ K. 
The calculated lattice parameter is 7.62 {\AA} and the corresponding band structure is shown in Fig. \ref{fig:1Lunbands}, 
describing a semiconducting system with a spin-up channel band gap $E_{g, \uparrow} =$ 0.552 eV and
a spin-down channel band gap $E_{g, \downarrow} =$ 1.273 eV
(highlighted by the shaded orange regions in the figure).
It should be pointed out that other \pcom{and more complex} magnetic states, arising in larger (e.g. 2$\times$2) supercells,
can be devised, at an energy from one to few tens of meV
higher than the FM ground state \cite{Nb3I8_PRR}. This is an interesting point because, despite the FM state is
definitely stable over a wide range of
temperatures, phase transitions between different magnetic states could easily occur under suitable conditions
(e.g. applied magnetic fields).

\pcom{The dynamical stability of Nb$_3$I$_8$-1L has been demonstrated both
experimentally (isolated monolayers have been successfully obtained onto suitable substrate \cite{https://doi.org/10.1002/pssr.201800448})
and theoretically (from the calculated phonon spectrum, as obtained within the same theoretical framework
as that of the present work \cite{C6NR07231C}).}
It should be pointed out that, as far as the multilayer and the strained
nanofilms considered in the next are concerned, the assessment of the dynamical stability has not be carried out, being computationally quite demanding and out of the scope of the present work. Here, we are going to conceptually support, with the presented results, how the interplay between magnetism, stacking and strain makes Nb$_3$I$_8$ an ideal candidate for novel magnetic devices and functionalities, that certainly demand for further work and investigation.

\gcom{To conclude the discussion on Nb$_3$I$_8$-1L, the in-plane magnetic coupling, leading to the
discussed FM magnetic ordering, can be estimated by building a $2\times2$ supercell, as difference
between the FM ground state energy and the energy obtained after flipping just one of the spins,
out of the four available. The magnetic coupling so obtained is about 23 meV. Such result seems
independent on the size of the supercell, since the same value is obtained for a
 $3\times3$ supercell.}

\subsubsection{\pcom{Stacking}}
Starting from Nb$_3$I$_8$-1L, \pcom{the next layer} with bulk stacking is 
obtained by \pcom{applying to the former suitable in-plane coordinate} transformations.
For instance, the second layer results from \pcom{applying a spatial} inversion operation \dcom{of the first one}
\pcom{with respect to the center of an Nb$_3$ cluster} (inversion center),
\pcom{whereas the third layer results from the first one after applying a suitable fractional translation.}
\pcom{On the other hand, multilayer nanofilms with AA stacking can be easily obtained by just replicating
the first layer at suitable distances, at fixed in-plane coordinates.}

By assuming the (most stable) FM spin ordering within a single layer, the question arises on whether,
after stacking two or more layers on top of each other with a given stacking geometry, consecutive layers
preferentially carry out the same spin or opposite spins (thus giving rise to a ``mixed'' spin configuration,
with all triangular Nb$_3$ clusters in the same plane exhibiting parallel spins,
but with a spin flip when moving form one layer to a nearby one).
This is a key concept because the possibility of obtaining a ``layered''
and eventually tunable (by means of electrostatic doping or out-of-plane pressure) magnetism paves the way
to a wide range of applications\dcom{, as already
inferred for CrI$_3$} \cite{Jiang2018, Huang2018}.

\begin{figure*}
\includegraphics[scale=0.4]{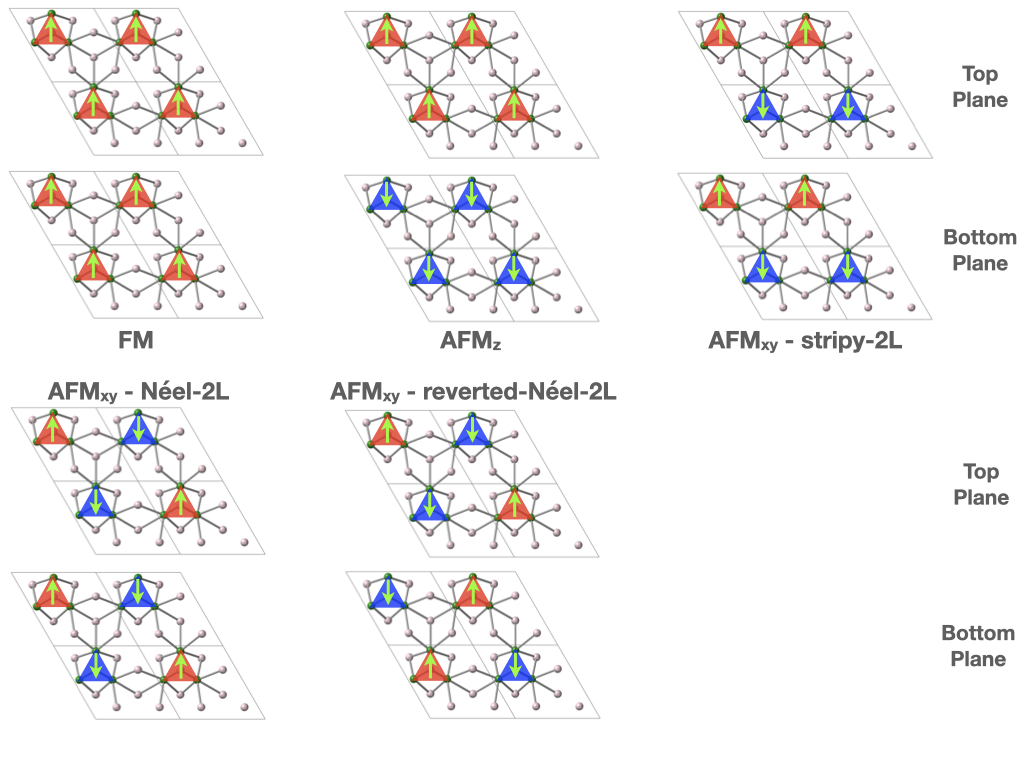}
\caption{ \pcom{$2\times2$ supercell and spin patterns investigated for Nb$_3$I$_8$-2L. The naming follows
the same convention as in the text.
Nb$_3$ clusters are highlighted with triangles. Red (blue) triangle background refers to spin-up (spin-down),
also depicted with up (down) green arrows. It is evident that FM and AFM$_z$ are characterized by a non-zero
magnetization of each Nb$_3$I$_8$ single layer whereas the total magnetization is not zero only for FM.} }
    \label{fig:afmstates}
\end{figure*}

\subsubsection{\pcom{Nb$_3$I$_8$-2L}}
\pcom{For Nb$_3$I$_8$-2L we explored }
both bulk and AA stackings\pcom{, that reveal an in-plane optimized lattice parameter
 quite similar to that of Nb$_3$I$_8$-1L ($\simeq$ 7.62 {\AA}).}

\pcom{As far as the magnetic ordering is concerned, the most straightforward configuration to be conceived
is that of stacking the two layers (with either bulk or AA stacking) with in-plane FM ordering, same as that
of Nb$_3$I$_8$-1L (see Fig. \ref{fig:structure}(c)).
In this respect, each layer becomes an ``elementary'' building block of the multilayer film,
carrying a $\pm1\mu_B$ magnetic moment per unit cell
(all Nb$_3$ clusters belonging to the same plane carry the same spin).
The remaining degree of freedom is that of the relative sign of such moment between
the two layers. Therefore, we can conceive two
different out-of plane orderings, that will be referred to as 
FM or $\uparrow\uparrow$ 
(where both layers have a $+1 \mu_B$ magnetic moment and
and the total magnetization is non-zero as well) and
AFM$_z$ or $\uparrow\downarrow$ (where the two layers carry opposite spins, each layer as a non-zero
magnetization whereas the total magnetization is zero).}

\pcom{However, aimed at giving a more comprehensive picture of the possible magnetic patterns
and their relative stability,
$2\times2$ supercells were also employed, to explore configurations where each single layer
carries a zero total magnetization (two out of the four Nb$_3$ clusters in each plane carry a spin up,
the other two a spin down, each plane has a zero magnetization and the
total magnetization is zero as well). These configurations
will be referred to as AFM$_{xy}$ and are compatibles with
different spin patterns, named ``stripy-2L'', ``N\'eel-2L'', and 
``reverted-N\'eel-2L'' states. These patterns, together with FM and AFM$_z$, are
sketched in} Fig. \ref{fig:afmstates}.
In the stripy-2L \pcom{ordering, spin up and spin down in each plane are distributed according to alternating rows,
and the same pattern is identically repeated in the two layers. 
In the N\'eel-2L ordering  spin up and spin down in each plane are distributed according
to the supercell diagonal (that is, Nb$_3$ clusters belonging to the same diagonal carry the same spin),
and the same pattern is identically repeated on the two layers.
Finally, the reverted-N\'eel-2L state is obtained by flipping the spins of
the N\'eel-2L state in the second layer,
such that to a spin-up in the
top layer corresponds a spin-down in the bottom layer and the other way around for spin-down.
The spin densities associated with the AFM$_{xy}$ orderings are reported
in Sec. I of the SM for the sake of completeness.}

\pcom{In the following we report the analysis of the different FM, AFM$_z$ and AFM$_{xy}$ magnetic orderings.}
\pcom{The relative energy of each magnetic pattern with respect to the lowest-energy AFM$_z$
ordering is
reported in Tab. \ref{tab:2Lunstrained} for both AA and bulk stacking.
The energies are reported per $1\times1$ unit cell (u.c.),
so as to allow a straightforward comparison between $1\times1$  and $2\times2$ supercells. }
It turns out that, regardless of the stacking geometry, 
the magnetic ground state definitely turns out to be much more stable than the NM solution and
\pcom{that AFM$_z$, among the considered ones, is always
the lowest-energy spin pattern .
Such circumstance at least partly
distinguishes Nb$_3$I$_8$ from CrI$_3$. Indeed, as far as the latter is concerned,
while a similar interlayer AFM$_z$ ordering has been argued
for the bilayer, a stacking-dependent magnetism shows
up~\cite{PhysRevResearch.3.013027,doi:10.1021/acs.nanolett.8b03321,doi:10.1021/acs.nanolett.1c02096}, that seems
to lack in Nb$_3$I$_8$. However, in the case of Nb$_3$I$_8$
the stacking geometry plays a role in the relative stability of
different magnetic states.
Indeed, AFM$_{xy}$ orderings, as depicted in  Fig. \ref{fig:afmstates}, depending on the stacking,
show higher energies, ranging from about 10-20 meV to about 100 meV. On the other hand,
the FM state energy is $\sim$77 meV/u.c. and $\sim$6 meV/u.c. higher than AFM$_z$
in the bulk and the AA stacking,
respectively.
These results reveal that an important role might be played by the deposition steps in the
fabrication of real samples, in that the stability of the
AFM$_z$ ordering against other magnetic orderings gets much more pronounced
for bulk stacking.}
As such, the magnetic phase diagram and its dependence on the temperature can be modified by
effect of the stacking geometry.

\begin{figure*}
(a)\includegraphics[scale=0.28]{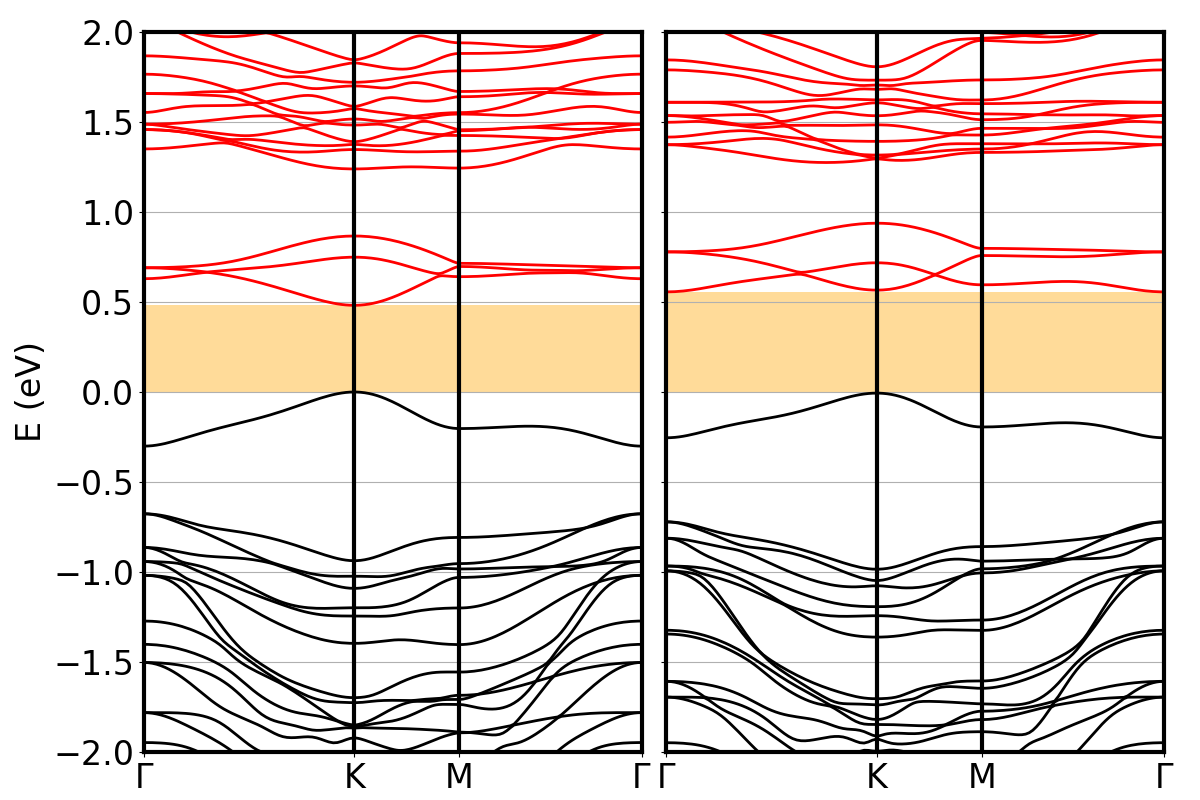}\includegraphics[scale=0.28]{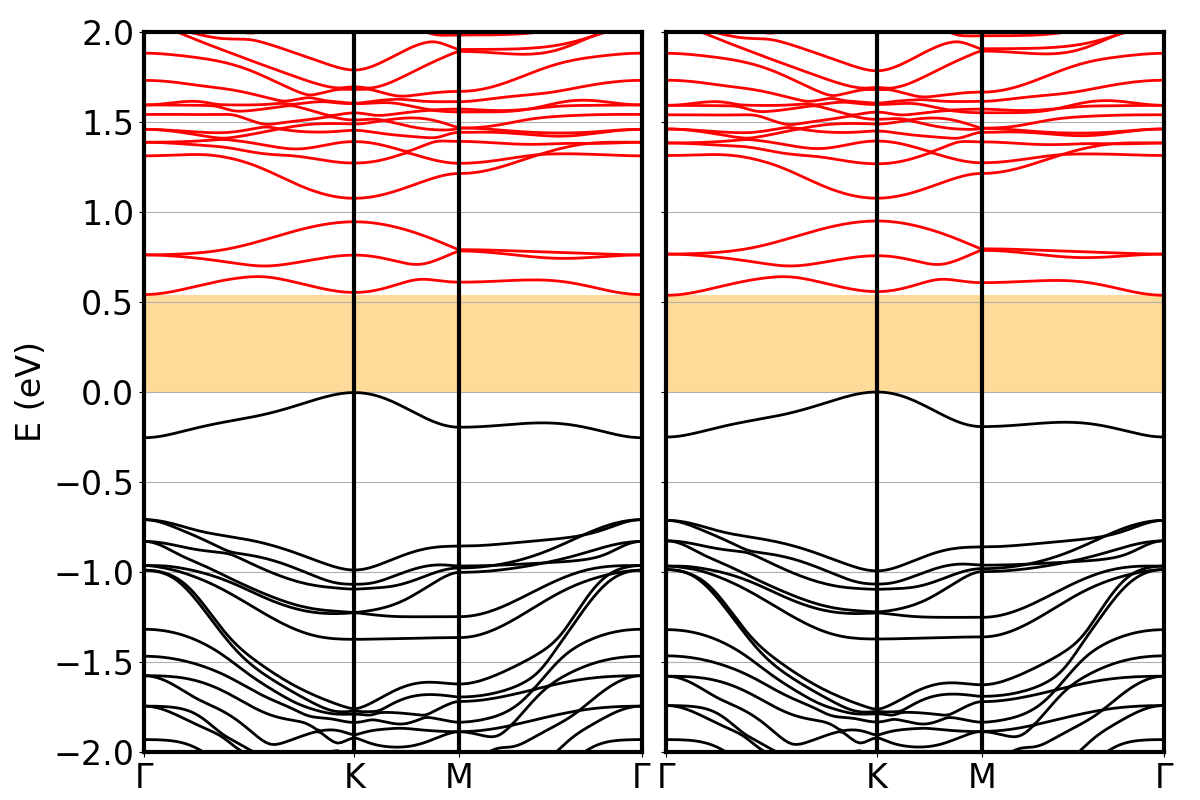}(b)
    \caption{Spin-up (left panels) and spin-down (right panels) band structure  along
    the $\Gamma-K-M-\Gamma$ path in the first BZ 
    of Nb$_3$I$_8$-2L (in the lowest-energy magnetic configuration\pcom{, AFM$_z$})
    for (a) AA and (b) bulk stacking. Zero energy corresponds to the top valence band. The band structure shows a semiconducting behavior, with the energy gap highlighted, for each spin channel, by a shaded orange region. }
    \label{fig:2Lunbands}
\end{figure*}

The effect of the stacking geometry also emerges from the analysis of the spin-polarized band structure.  In Fig. \ref{fig:2Lunbands},
the band structure for both AA and bulk stacking is shown for the lowest-energy AFM$_z$ state. 
It can be clearly inferred that for bulk stacking spin-up and spin-down channels provide almost identical band structures (see Fig. \ref{fig:2Lunbands}(b)), 
with a very tiny difference between the corresponding gaps, $E_{g, \uparrow}=0.544$ eV and $E_{g, \uparrow}=0.537$ eV. This can be ascribed to the 
inversion symmetry \pcom{linking the top and the bottom layer in} the bulk stacking.
On the other hand, for AA stacking, although the overall band structures look similar, 
the breaking of the inversion symmetry results in a $k$-point dependent spin splitting, yielding the band gaps $E_{g, \uparrow}=0.481$ eV and 
$E_{g, \downarrow}=0.562$ eV (see Fig. \ref{fig:2Lunbands}(a)).
\pcom{The presence or lack of inversion symmetry for the two stackings
can also be easily identified in the
charge transfer plots, reported in Sec. I of the SM and showing the ground-state
electronic charge difference between Nb$_3$I$_8$-2L and the isolated top and bottom layer.}

\begin{table}[]
    \centering
    \begin{tabular}{c|c|c}
    \hline
    \multirow{2}{*}{Magnetic state}  &   \multicolumn{2}{c}{$\Delta E$ (meV/u.c.)}  \\
                    &   AA stacking   &   bulk stacking   \\
    \hline
    NM              &   292.1           &   103.0           \\
    FM \pcom{($\uparrow\uparrow$)}              &   5.6             &   76.7            \\
    AFM$_{xy}$-N\'eel-2L            &   21.2            &   96.2            \\
    AFM$_{xy}$-reverted-N\'eel-2L   &   15.6            &   13.7            \\
    AFM$_{xy}$-stripy-2L          &   21.2            &   96.2
    \end{tabular}
    \caption{Relative stability for different magnetic states of Nb$_3$I$_8$-2L for both bulk and AA stacking.
    $\Delta E$ is the energy difference between the considered state and 
    the lowest-energy one, that \pcom{for both stackings} corresponds to the AFM$_z$
    \pcom{($\uparrow\downarrow$).} \dcom{state, i.e., the in-plane FM magnetic ordering with AFM}
\dcom{ordering between the top and bottom planes.} NM stands for nonmagnetic state, FM for ferromagnetic state (both layers carry the same spin\pcom{, $\uparrow\uparrow$}), whereas
    N\'eel-2L, reverted-N\'eel-2L, and stripy-2L states are as described in the text. The energies are calculated per \pcom{$1\times1$ unit cell (u.c.)}.}
    \label{tab:2Lunstrained}
\end{table}

\subsubsection{\pcom{Nb$_3$I$_8$-3L}}
At this stage, one might wonder about what would happen if thicker films with an odd number
of layers would be considered.
Since the AFM$_z$ magnetic ordering\pcom{, corresponding to AFM coupling between two
consecutive layers,} was proven to be the most stable for Nb$_3$I$_8$-2L,
we analyzed similar patterns for Nb$_3$I$_8$-3L. \pcom{Given the in-plane FM ordering in each
layer,}
different out-of-plane spin-stacking sequences were considered: $\uparrow\uparrow\uparrow$,
analogous of the FM
state of Nb$_3$I$_8$-2L; $\uparrow\downarrow\uparrow$, where neighbor layers carry
opposite spin and obtained from the first by flipping the central layer spin;
$\uparrow\uparrow\downarrow$, obtained from the first
by flipping the spin of one of the outermost layers.
The optimized lattice parameters are 7.63 {\AA} and 7.62 {\AA} for the bulk and
the AA stacking, respectively.
The \pcom{$\uparrow\downarrow\uparrow$ ordering, corresponding to AFM coupling between each pair
of consecutive layers,} has \pcom{been} proven to be the most stable configuration,
in agreement with what already reported for trilayer CrI$_3$ \cite{Song2019}.
In Tab. \ref{tab:3Lunstrained}, we report the relative energies per u.c. of all the 
considered magnetic states with respect to the \pcom{lowest-energy} $\uparrow\downarrow\uparrow$ ordering,
for both bulk and AA stacking.
Again, the stacking geometry reveals its central role, in that for bulk stacking several tens
of meV separate $\uparrow\uparrow\uparrow$
and $\uparrow\uparrow\downarrow$ from the ground state.
On the other hand, they lie only $\simeq$12 and $\simeq$6 meV from the ground state in the AA stacking.

\begin{figure*}
(a)\includegraphics[scale=0.28]{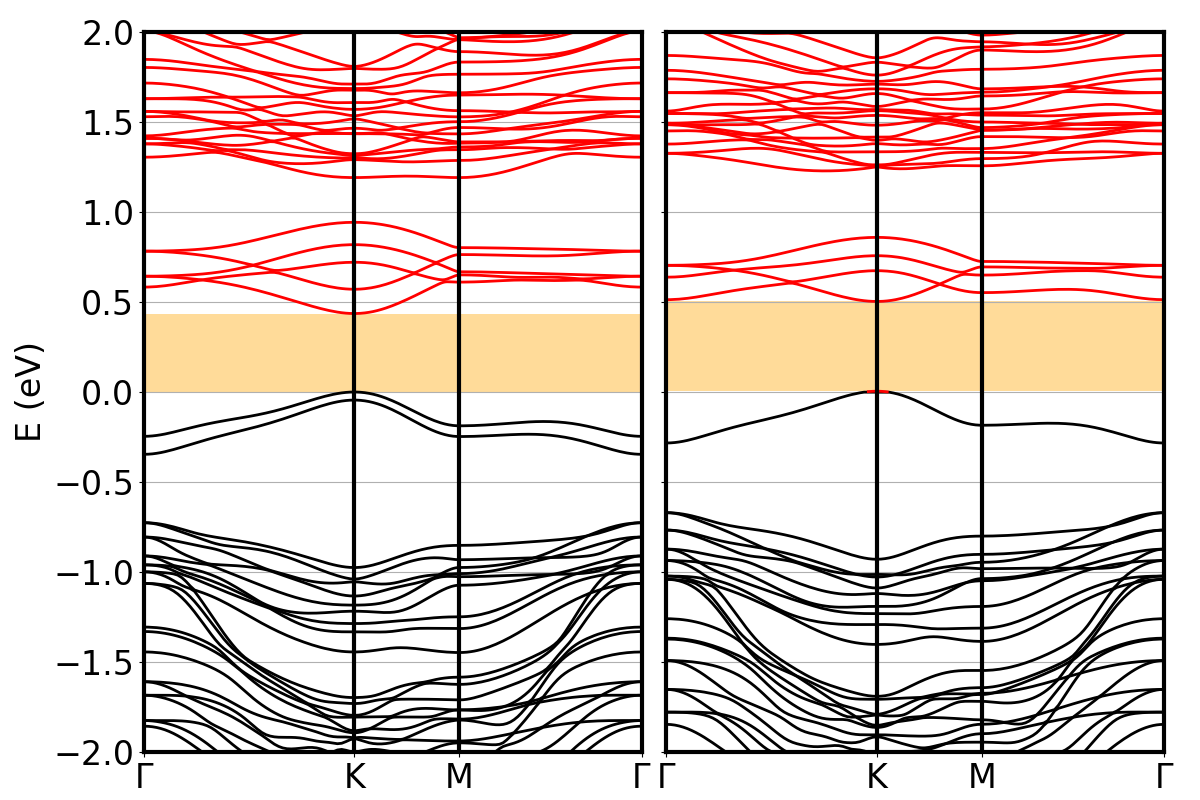}\includegraphics[scale=0.28]{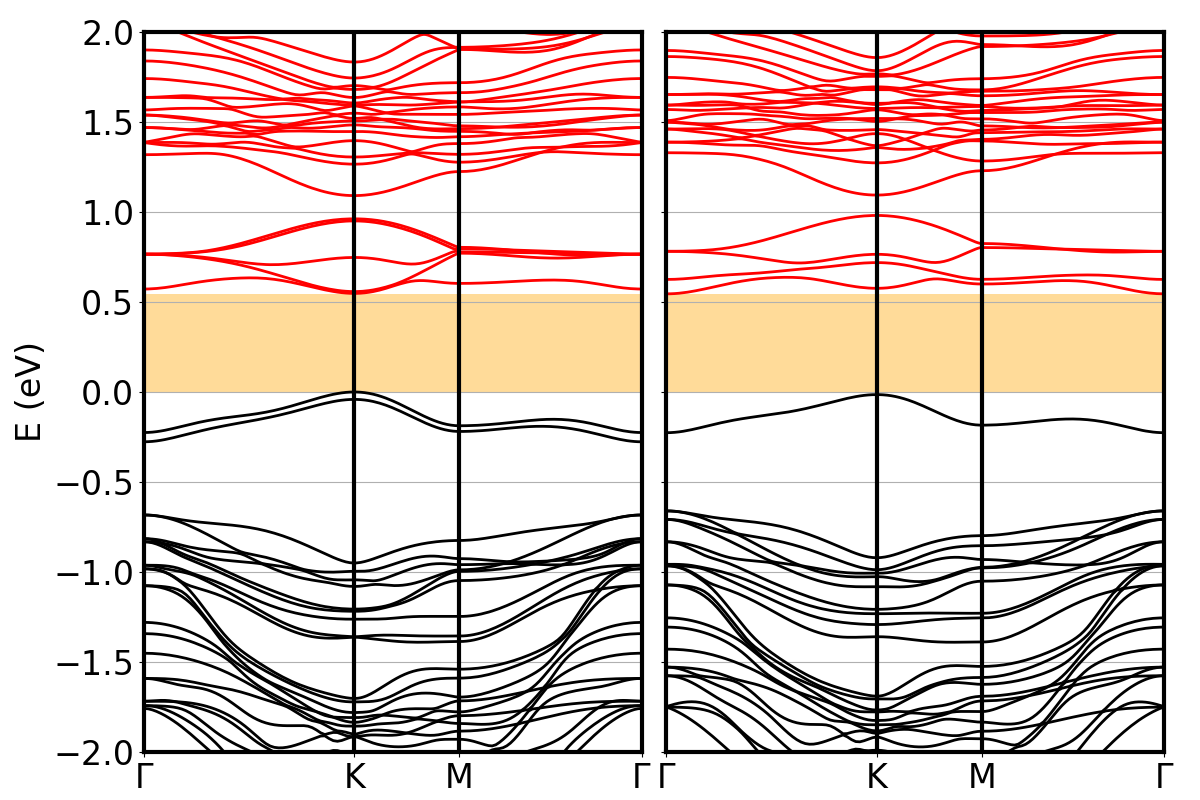}(b)
    \caption{Spin-up (left panels) and spin-down (right panels) band structure along
    the $\Gamma-K-M-\Gamma$ path in the first BZ 
    of Nb$_3$I$_8$-3L (in the lowest-energy magnetic configuration, $\uparrow\downarrow\uparrow$)
    for (a) AA and (b) bulk stacking. Zero energy corresponds to the top valence band.
    The band structure shows a semiconducting behavior, with the energy gap highlighted,
    for each spin channel, by a shaded orange region.}
    \label{fig:3Lunbands}
\end{figure*}

As far as the ground-state spin-polarized band structure is concerned, we do not expect any degeneracy for both stacking geometries, 
since no inversion symmetry operation can be identified. 
Indeed, the bulk stacking shows a semiconducting band structure for both spin channels with $E_{g, \uparrow}=0.547$ eV and 
$E_{g, \downarrow}=0.559$ eV (see Fig. \ref{fig:3Lunbands}(a)), whereas the AA stacking provides $E_{g, \uparrow}=0.436$ eV 
and $E_{g, \downarrow}=0.498$ eV (see Fig. \ref{fig:3Lunbands}(b)).

\begin{table}[]
    \centering
    \begin{tabular}{c|c|c}
    \hline
    \multirow{2}{*}{Magnetic state}  &   \multicolumn{2}{c}{$\Delta E$ (meV/u.c.)}  \\
                    &   AA stacking   &   bulk stacking   \\
    \hline
    NM              &   422.7           &   276.7           \\
    $\uparrow\uparrow\uparrow$              &   11.9             &   79.5            \\
    $\uparrow\uparrow\downarrow$          &   5.9            &   78.0
    \end{tabular}
    \caption{Relative stability for different magnetic states of Nb$_3$I$_8$-3L for both bulk and AA stacking. 
    $\Delta E$ is the energy difference per formula unit with respect to the lowest-energy state, that in both cases corresponds to the $\uparrow\downarrow\uparrow$ (AFM) state, i.e., the FM magnetic ordering with AFM ordering between the top, middle and bottom plane. $\uparrow\uparrow\uparrow$ stands for the (ferromagnetic) ordering where all planes carry out the same spin and $\uparrow\uparrow\downarrow$ for the ordering where two consecutive planes carry the same spin, opposite to that of the third plane.}
    \label{tab:3Lunstrained}
\end{table}

\pcom{As a final remark, we should point out that the out-of-plane AFM magnetic coupling can be
estimated, from our results, to be of the order of 6 meV for AA stacking and 80 meV in bulk stacking.
Such a coupling is calculated as the
the energy needed to flip the spin of a whole layer starting from an otherwise AFM$_z$
magnetic configuration and
can inferred from both the results for Nb$_3$I$_8$-2L
(as $E_{\uparrow\downarrow} - E_{\uparrow\uparrow}$ in Table \ref{tab:2Lunstrained})
and those for Nb$_3$I$_8$-3L (as
$E_{\downarrow\uparrow\downarrow} - E_{\uparrow\uparrow\downarrow}$
in Table \ref{tab:3Lunstrained} ). As such, we can conclude that, while the stacking
seems not to be able to modify the lowest-energy magnetic configuration, bulk stacking
can definitely make it more stable than it is in the AA stacking.}

\subsection{Strained \dcom{systems} \pcom{nanofilms} }
\label{ssec:strained}
\pcom{Applied strain can directly impact} on the electronic properties because it affects the
interactions between the atoms composing the lattice. 
In particular, since the magnetism arises from the exchange interactions between the magnetic
atoms belonging to the lattice, 
strain can be devised as an effective degree of freedom for tuning the magnetic couplings and,
as a consequence, move the system 
across the magnetic phase diagram in an absolutely unpredictable way. 
Here, we analyze in-plane compressive or tensile \pcom{biaxial} strain ranging from
$-7.5$\% to 7.5\% in terms of lattice constant variations and discuss
to what extent it can induce magnetic phase transitions.

%\begin{figure*}
%\hspace*{-1cm}(a)\hspace*{4.5cm}(b)\hspace*{4.5cm}(c)\vspace*{0.1cm}\\
%\includegraphics[scale=0.065]{Neel-1L.png}\includegraphics[scale=0.065]{stripy-1L.png}\includegraphics[scale=0.065]{Nb3-island.png} \vspace*{0.25cm}
%    \caption{AFM spin patterns in Nb$_3$I$_8$-1L: (a) ``stripy-1L'', (b) ``N\'eel-1L'', and (c) ``Nb$_3$-island-1L'' configurations. Yellow (blue) isosurfaces correspond to positive (negative) spin density.}
%    \label{fig:afmstates_strained-1L}
%\end{figure*}

\begin{figure}
\includegraphics[scale=0.35]{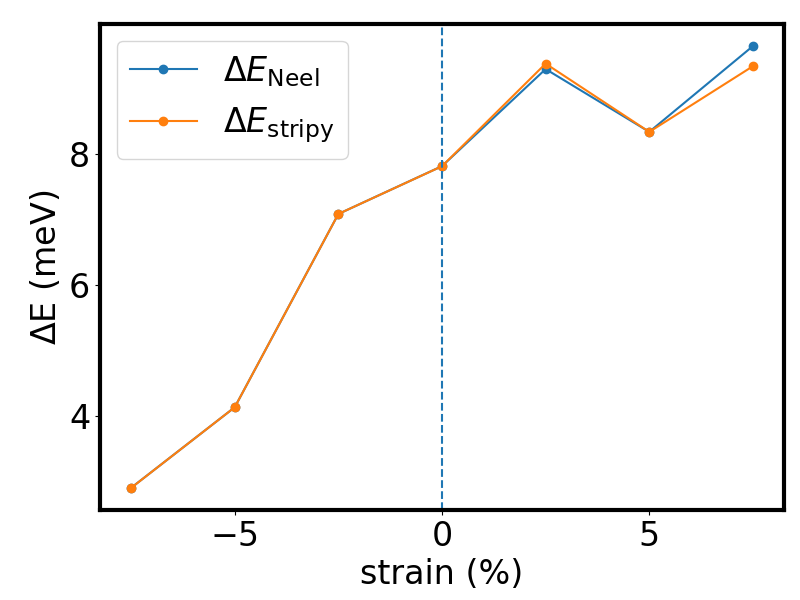}\vspace*{0.5cm}
    \caption{Relative stability of different magnetic orderings in 
    Nb$_3$I$_8$-1L. $\Delta E = E - E_\mathrm{FM}$ is the energy difference
    between the energy of a given configuration and that of the lowest
    energy, FM state: ``N\'eel-1L'' and ``stripy-1L'' configurations are considered.
    \pcom{NM state shows a similar trend, but is not shown because it lies at higher energy in
    all the considered range. The energies are calculated per $1\times1$ unit cell.} The very tiny difference
    between the two curves showing up at the largest positive strain ($\sim 0.3$ meV) is likely the result
    of numerical noise. }
    \label{fig:curves-1L}
\end{figure}

\subsubsection{\pcom{Nb$_3$I$_8$-1L}}
Let us start our analysis from Nb$_3$I$_8$-1L. \pcom{Here, two main questions arise: the first is about
whether the strain might stabilize other magnetic phases against the stable FM phase found for the unstrained
monolayer. Second, whether and to what extent the electronic properties of a given magnetic phase
may change as an effect of the strain.}

\pcom{Aimed at giving an answer to the first question, besides the NM and
FM orderings previously discussed for Nb$_3$I$_8$-1L, here we consider also, for the sake of completeness,
two other configurations, that is, AFM$_{xy}$-stripy-1L and AFM$_{xy}$-N\'eel-1L (with definitions similar to
those of Fig. \ref{fig:afmstates}, by keeping in mind that the figure
has been conceived for bilayers whereas here we are dealing with
a monolayer).}

%In addition to the NM, FM, AFM$_{xy}$-stripy-1L (see Fig. \ref{fig:afmstates_strained-1L}(b)) and AFM$_{xy}$-N\'eel-1L (see Fig. \ref{fig:afmstates_strained-1L}(a)) magnetic configurations
%already discussed for the unstrained systems, we also consider the Nb$_3$-island-1L configuration (see Fig. \ref{fig:afmstates_strained-1L}(c)) as a further FM 
%state, consisting of a single Nb$_3$ cluster having opposite spin compared to the other ones in a $2\times2$ supercell. 

\pcom{The FM magnetic ordering turns out to be stable against strain effects for
all the considered values of tensile and
compressive strain. However, larger compressive strain makes the different magnetic states closer in
energy to be contrasted with tensile strain that instead further stabilizes the FM state. This can be 
be easily inferred from Fig. \ref{fig:curves-1L},
where the relative energy of different magnetic configurations is
reported versus}
the applied strain. 
This is a striking result because it demonstrates the possibility of using strain as a control
knob to stabilize the magnetic phase during the deposition steps, since a sufficiently high
tensile (compressive) strain enhances (reduces) the energy difference between NM and FM states
\pcom{and between different magnetic states}.
\pcom{The explanation of such behavior requires to remind that, as previously stated, the origin
of magnetism in Nb$_3$I$_8$ stands in the irregular Kagom\'e lattice, where each  Nb$_3$ triangular cluster
carries a 1 $\mu_B$ magnetic moment, that is, a 1/2 spin. The magnetic interactions between those clusters
are modified by strain, in particular because an enhanced or reduced inter-cluster distance.}

%From the analysis of the spin densities, it is clear that the FM state is characterized by
%the magnetization of the Nb$_3$ triangular clusters, each carrying a
%1 $\mu_B$ magnetic moment, that is, a 1/2 spin.
%By identifying each cluster with its total 1/2 spin, such systems appears as a triangular
%lattice of 1/2 spins, whose stability can be tuned with strain.
%In particular, tensile strain enhances the stability of this phase against the others 
%(since cluster-cluster couplings that might induce other magnetic orderings are weakened because of the increased intercluster distance).

\begin{figure*}
(a)\includegraphics[scale=0.28]{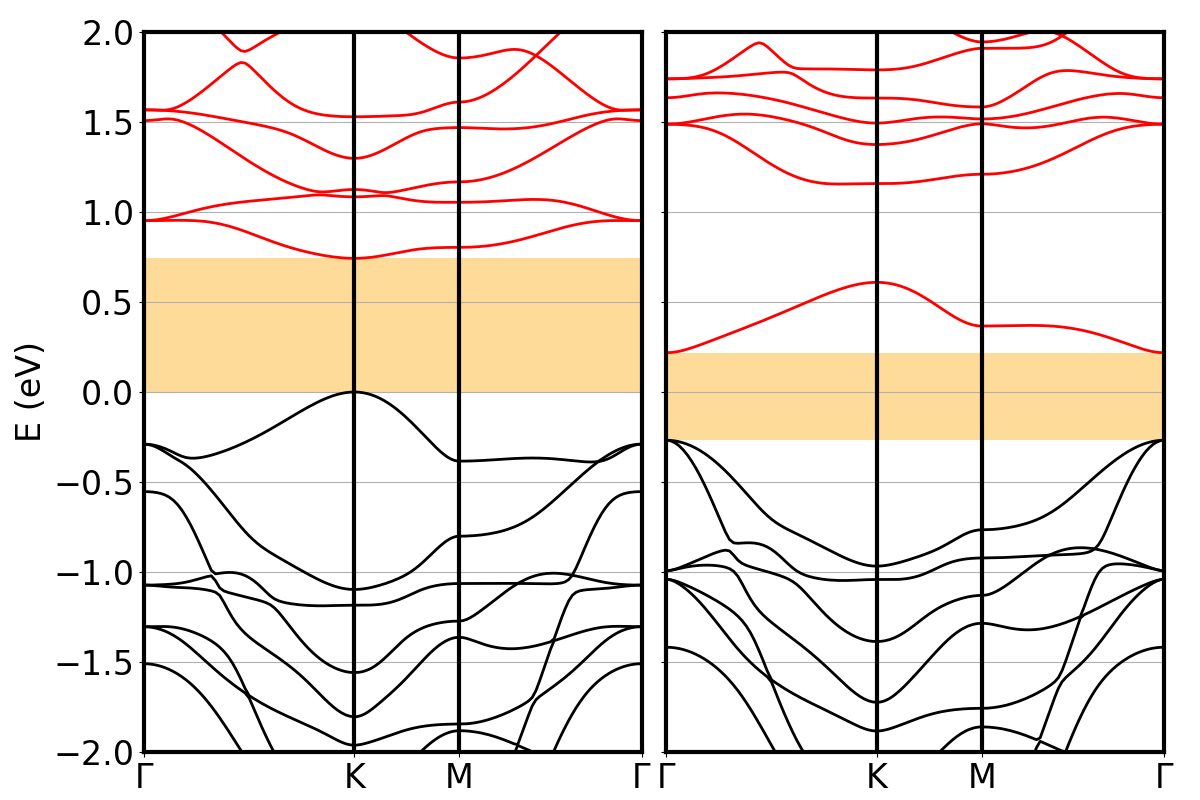}\includegraphics[scale=0.28]{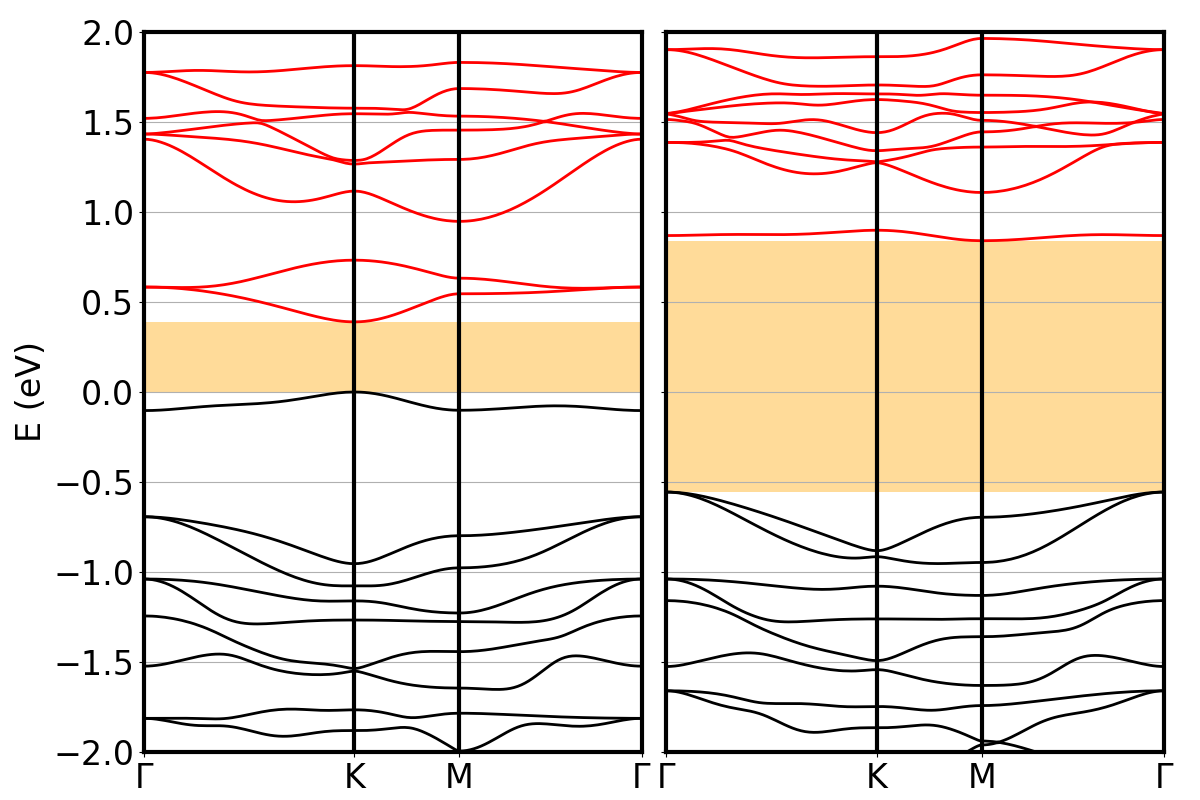}(b)\vspace*{0.5cm}
(c)\includegraphics[scale=0.35]{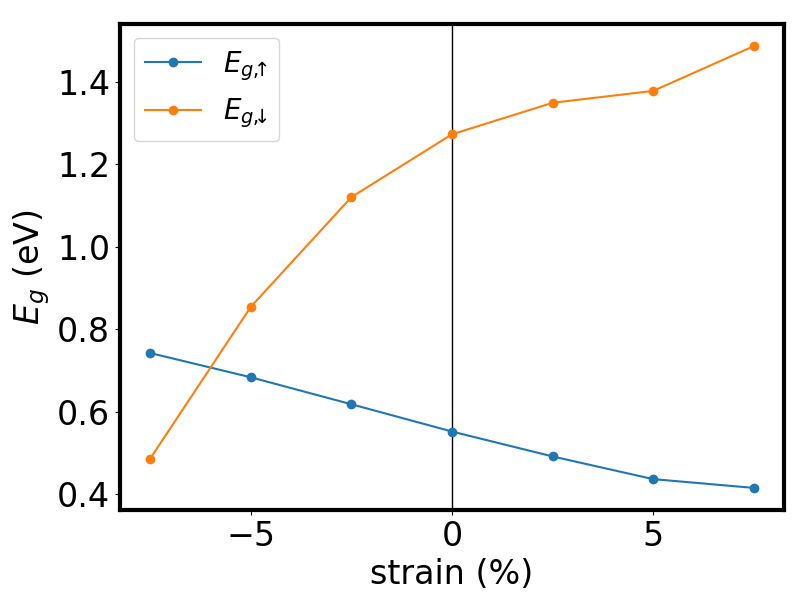}
    \caption{Spin-up (left panels) and spin-down (right panels) band structure along
    the $\Gamma-K-M-\Gamma$ path in the first BZ 
    of Nb$_3$I$_8$-1L (in the lowest-energy magnetic configuration\pcom{, FM}) under a (a) $-7.5$\% and (b) 7.5\% strain. Zero energy corresponds to the top valence band. The band structure shows a semiconducting behavior, with the energy gap highlighted, for each spin channel, by a shaded orange region.
    \pcom{(c) Spin-up ($E_{g,\uparrow}$) and spin-down ($E_{g,\downarrow}$)
    energy gap of Nb$_3$I$_8$-1L (in the lowest-energy FM 
    configuration) as a function of the strain.
    We can notice the decreasing (increasing) behavior of $E_{g,\uparrow}$ ($E_{g,\uparrow}$) 
    when tuning the strain from compressive to tensile.}}
    \label{fig:1Lbands}
\end{figure*}

\pcom{Now we turn to the second question, that was about to strain-induced effects on the
electronic properties. To this aim, in}
Fig. \ref{fig:1Lbands} \pcom{the spin-up and spin-down}
band structures of Nb$_3$I$_8$-1L 
in the presence of a $\pm$7.5\% strain are reported.
\pcom{As a general remark}, in the strong tensile strain regime,
the electronic bands exhibit a flattening deriving 
from the increas\pcom{ed} in-plane interatomic distances.
A \pcom{more intriguing}  effect of the strain on the band structure concerns the 
band gap variation: by tuning the strain from compressive to tensile,
a decrease (increase) of the spin-up (spin-down) channel band gap 
is observed, \pcom{as we show in Fig. \ref{fig:1Lbands}(c).}
Such an opposite behavior for spin-up and spin-down channel may be ascribed to the number and nature of
electronic bands in proximity of the energy gap region. As it will be further clarified
in a while from the projection of the energy bands onto atomic orbitals, the relative
\pcom{contribution} of I and Nb orbitals to those bands is
differently influenced by the strain for the two spins, with a direct effect onto the 
energy gaps. We could expect that this intriguing peculiarity might be unveiled from absorption experiments, as an example.

An in-depth analysis also shows that upon increasing strain a decreasing 
of the monolayer thickness is observed, ranging from $\sim4.57$ {\AA} for a $-7.5$ \% strain to $\sim3.67$ {\AA} for a $7.5$ \% strain, to be compared with 4.08 {\AA} of the unstrained bilayer
(that is indeed intermediate
between the other two). In other words,
tensile (compressive) strain tends to weaken (enhance) the interatomic interaction along the $z$ direction.

\begin{figure}
\includegraphics[scale=0.28]{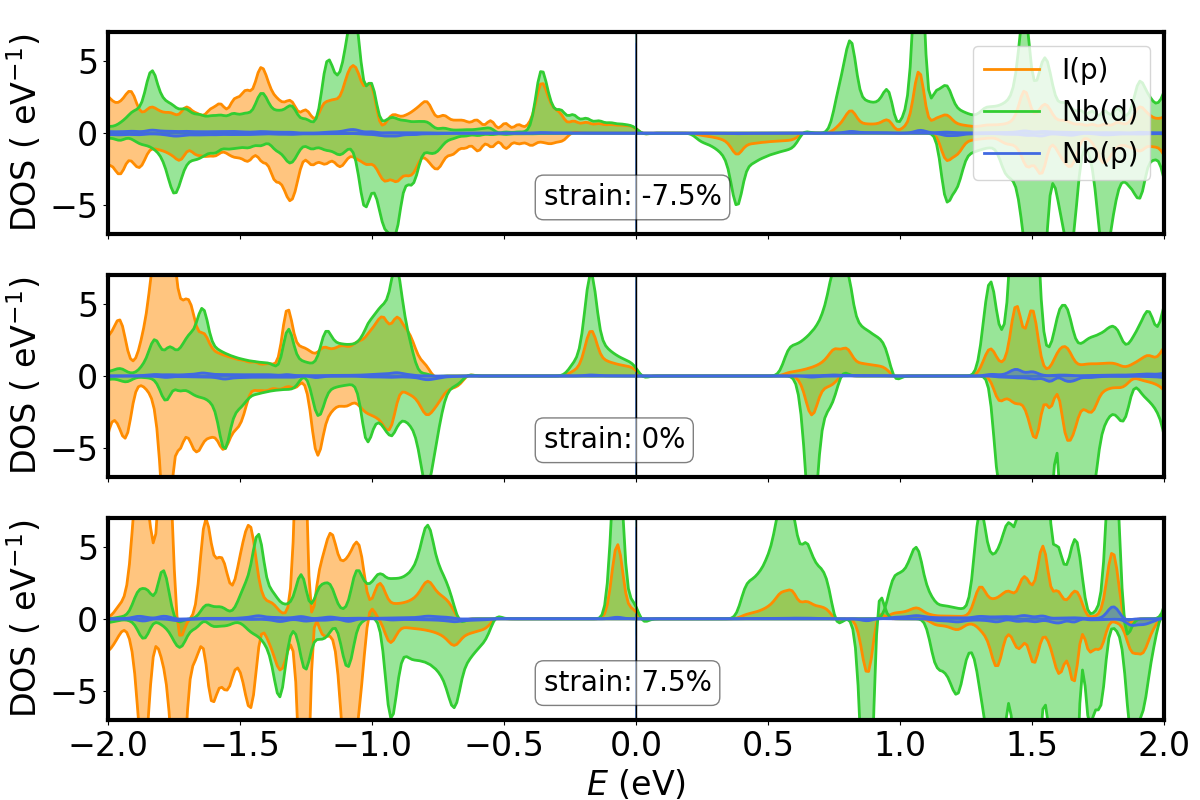}
    \caption{PDOS of Nb$_3$I$_8$-1L in the lowest-energy FM configuration onto I($p$), Nb($d$) and Nb($p$) 
    orbitals. The largest considered compressive ($-7.5$\%) and tensile 
    (7.5\%) strains, together with the unstrained system are considered.
    Zero energy corresponds, for each system, to the top valence band.
    Positive (negative) values of the PDOS correspond to spin-up
    (spin-down) bands.}
    \label{fig:PDOS}
\end{figure}

\pcom{As a final remark, we would like to point out that strain-induced effect on the electronic
structure can also be related to the change in the orbital hybridization following structural modification.
As an example, we report in Fig.  \ref{fig:PDOS}}
the projected density of states (PDOS) onto atomic orbitals\pcom{.
Almost independently of the strain}, the top valence bands are mostly dominated by
hybridized Nb($d$) and I($p$) orbitals, with the larger contribution 
coming from the former. However, by looking at an energy window $\sim$0.4 eV below the top valence band,
the integrated PDOS shows that, on going from the largest compressive strain to the unstrained system
to the largest tensile strain the I($p$) contribution to the total PDOS changes from 45\% to 32\% to
30\%, respectively. We can infer that the compressive strain can significantly enhance the
hybridization between
I and Nb orbitals, whereas the opposite effect is observed in the case of tensile strain.

\subsubsection{\pcom{Nb$_3$I$_8$-2L and Nb$_3$I$_8$-3L}}
\begin{figure}
\includegraphics[scale=0.4]{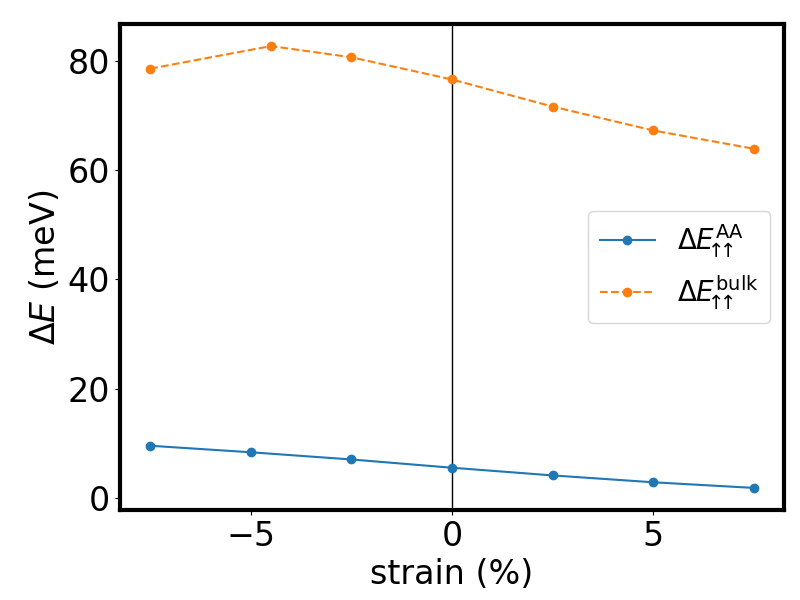}
    \caption{Relative stability of the $\uparrow\uparrow$ magnetic ordering in 
    Nb$_3$I$_8$-2L for both AA and bulk stacking as a function of the strain, referred to that of the
    lowest-energy $\uparrow\downarrow$ state ($\Delta E = E_{\uparrow\uparrow} - E_{\uparrow\downarrow}$).}
    \label{fig:curves-2L}   
\end{figure}

\begin{figure}
\includegraphics[scale=0.4]{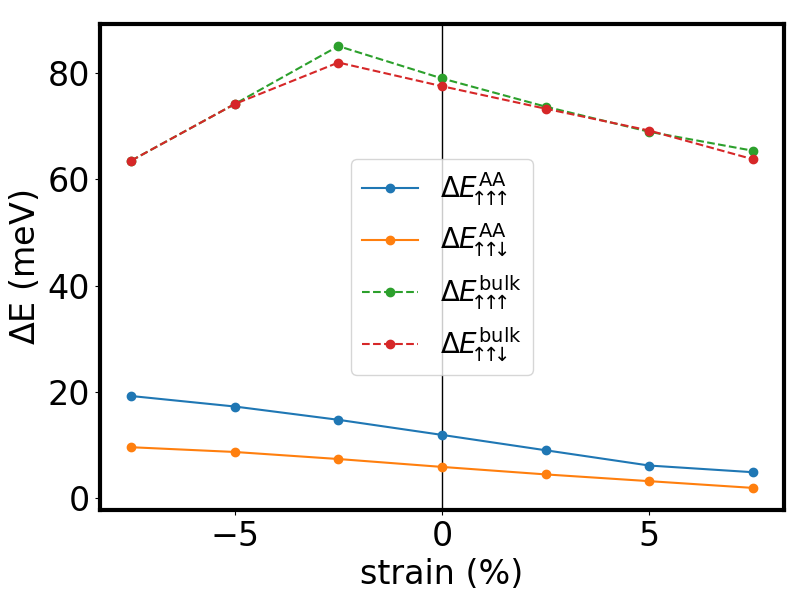}
    \caption{Relative stability of different magnetic orderings in 
    Nb$_3$I$_8$-3L for both AA and bulk stacking as a function of the strain.
    $\Delta E = E - E_{\uparrow\downarrow\uparrow}$ is the energy difference
    between the energy of a given configuration and that of the 
    lowest-energy, $\uparrow\downarrow\uparrow$ state.}
    \label{fig:curves-3L}
\end{figure}

Moving to the \pcom{thicker nanofilms,
in Figs. \ref{fig:curves-2L} and \ref{fig:curves-3L}, the energy differences 
of different magnetic orderings relative to the lowest-energy state
are reported, as a function of the applied strain and for both bulk 
and AA stackings. It is shown that strain does not significantly alter the 
ground-state magnetic ordering, in that 
%, with exception of Nb$_3$I$_8$-2L
%at large compressive strain, as detailed below.
adjacent layers
always carry opposite spin in the lowest-energy state 
($\uparrow\downarrow$ for Nb$_3$I$_8$-2L and $\uparrow\downarrow\uparrow$ 
for Nb$_3$I$_8$-3L).}

%\begin{figure}
%\includegraphics[scale=0.4]{Nb_mag_mom.png}
    %\caption{Magnetization of Nb$_3$I$_8$-2L \pcom{in the bulk stacking}
    %as a function of the applied
    %strain in the lowest-energy $\uparrow\downarrow$ state. The total magnetization is zero for all strains, the bottom and top layer contributions are shown.}
    %\label{fig:Nbmagmom}
%\end{figure}

%A particularly important effect is observed for Nb$_3$I$_8$-2L in the bulk stacking: at -7.5 \% strain and
%below, a phase transition to a NM state occurs, since Nb magnetic moments
%monotonically vanish, as illustrated in Fig. \ref{fig:Nbmagmom}. This does
%not hold for the AA stacking, providing further proof of the deposition
%steps importance for the expected magnetic state of the system.

\pcom{As previously reported for Nb$_3$I$_8$-1L,
strain-induced non-trivial modifications of the electronic structure
can be observed, in particular as far as spin-up and spin-down band gaps
are concerned and more evident for bulk stacking (this is
shown in Figs. S2 and S3 of the SM
for Nb$_3$I$_8$-2L and Nb$_3$I$_8$-3L, respectively).
Moreover, similarly to the Nb$_3$I$_8$-1L, band flattening
occurs in the strong tensile strain regime.}
%, since the system cannot arbitrarily compress along the $z$-axis direction because of the out-of-plane vdW interaction weakly gluing the layers. Thus, Nb $d_{z^2}$-orbitals keep on to offer electronic dispersion in the lattice.

\section{Conclusions}
\label{sec:IV}
This work explores the novel and intriguing magnetic and electronic
properties of few-layer Nb$_3$I$_8$ vdW nanofilms, showing how and to what
extent the magnetism and magnetic ordering can be considered as tunable
properties as a function of the film thickness, applied strain,
stacking geometry and the combined effect of these three parameters.
By considering one-, two-, and three-layer systems, we have shown different
magnetic patterns that might be energetically favored among the many
possible. Spin densities \pcom{are shown to be} mostly localized on 
\pcom{the Nb$_3$ triangular clusters natively present into the irregular Kagom\'e lattice}.
In particular, we argued that within each plane a FM ordering
is energetically favored, with parallel spins on all Nb atomic sites. \pcom{In this respect,
a single Nb$_3$I$_8$  layer behaves as a ``macroscopic'' spin carrying a 1 $\mu_B$ magnetic
moment per unit cell.}
However, as one or more layer are \pcom{stacked on the first, the inter-layer
magnetic interaction drives the system to an out-of-plane AFM ordering
of the spins, such that
two consecutive layers carry
opposite magnetizations}. On the other hand, other, both in-plane and out-of-plane
magnetic patterns
have been shown to be \pcom{feasible, with energies depending on the stacking geometry and
the nanofilm thickness.
In particular, the out-of-plane AFM magnetic coupling results more (less)
stable in the presence of bulk (AA) stacking, demonstrating how stacking might play a fundamental
role for designing new magnetic materials and devices with given functionalities.}

%We explored both AA and bulk stacking, showing that the stacking geometry
%might significantly affect the electronic and magnetic properties. 
Similarly, strain effects have been \pcom{evidenced}, from band flattening
taking place for sufficiently large tensile stress, to more dispersed bands with increasing
contribution from I orbitals for large modulus compressive stress.

Interestingly,
%a transition to a NM state shows up when bulk-stacked two-layer
%systems under $-7.5$\% strain are considered.
%Moreover,
the
strain is also capable of enhancing or weakening (according to its sign)
the energy differences between different possible magnetic orderings,
paving the way to a strain-tunable magnetic response.

\gcom{The magnetic coupling, responsible for the in-plane FM and the
out-of-plane AFM orderings has been estimated from total energy differences.
It turns out that it is required an energy of about 23 meV to flip a single spin
within a single FM layer. On the other hand, about
6 meV for AA stacking and 80 meV for bulk stacking per unit cell
are required to flip the spin of a whole layer starting from two antiferromagnetically
coupled layers.}

All these results\pcom{, also combined with recent experimental and theoretical outcomes
on the bulk counterpart \cite{arxiv.2203.10547},} shed light on new and intriguing properties
of this novel material, bringing it among the possible candidates to implement more
complex magnetic responses, in user-designed homo- and hetero-structures. \pcom{Our outcomes
represent a step forward in the search of thickness-dependent and strain-tunable
magnetism in 2D van der Waals materials, which is currently object of intense and
ongoing research \cite{Ci2022}.}

\pcom{Future work could involve the study of defected Nb$_3$I$_8$ vdW nanofilms,
with the aim of bringing out, depending on the defect nature and concentration,
the twofold role that might be played, especially in a magnetic system, by the impurities:
on one hand, unwanted defects might at least partially destroy the desired
(e.g. magnetic) properties. On the other hand, intentionally induced impurities
might represent a novel degree of freedom to enhance those
properties~\cite{doi:10.1021/nl901557x,doi:10.1063/5.0039979,https://doi.org/10.1002/sstr.202100077}.
Moreover, a thorough study of valley polarization and its interplay with strain would be
desirable, as being investigated in other classes of two-dimensional
materials~\cite{Zhang_2017,C9CP02404B,doi:10.1021/acsami.0c13988}.
}

%\section*{Acknowledgments}
\begin{acknowledgments}
We acknowledge the CINECA awards HP10CSICON ``QUANTERA'' and HP10CZRR24 ``UNIVERSE'' under the ISCRA initiative for the availability of high performance computing resources and support.
Financial support and computational resources from MUR, PON “Ricerca e Innovazione 2014-2020”, under Grant No. PIR01\_00011 - (I.Bi.S.Co.) are acknowledged.
\end{acknowledgments}

G.C. and F.C. contributed equally to this work.

% SUPPLEMENTARY INFORMATION %%%%%%%%%%%%%%%%%%%%%%%%%%%%%%%%%%%%%%%%%%%%%%%%%%%%%%%%%%%%%%%%%%
%\appendix
%\section{Convergence tests}
%\label{sec:appendix}
%\subsection{Subsection1}
%\label{ssec:subsection1}
%%%%%%%%%%%%%%%%%%%%%%%%%%%%%%%%%%%%%%%%%%%%%%%% END OF SUPPLEMENTARY INFORMATION %%%%%%%%%%%%%%%%%%%%%%%%%%%%%%%

\section*{References}

\bibliography{article}

%merlin.mbs apsrev4-1.bst 2010-07-25 4.21a (PWD, AO, DPC) hacked
%Control: key (0)
%Control: author (72) initials jnrlst
%Control: editor formatted (1) identically to author
%Control: production of article title (-1) disabled
%Control: page (0) single
%Control: year (1) truncated
%Control: production of eprint (0) enabled
\begin{thebibliography}{69}%
\makeatletter
\providecommand \@ifxundefined [1]{%
 \@ifx{#1\undefined}
}%
\providecommand \@ifnum [1]{%
 \ifnum #1\expandafter \@firstoftwo
 \else \expandafter \@secondoftwo
 \fi
}%
\providecommand \@ifx [1]{%
 \ifx #1\expandafter \@firstoftwo
 \else \expandafter \@secondoftwo
 \fi
}%
\providecommand \natexlab [1]{#1}%
\providecommand \enquote  [1]{``#1''}%
\providecommand \bibnamefont  [1]{#1}%
\providecommand \bibfnamefont [1]{#1}%
\providecommand \citenamefont [1]{#1}%
\providecommand \href@noop [0]{\@secondoftwo}%
\providecommand \href [0]{\begingroup \@sanitize@url \@href}%
\providecommand \@href[1]{\@@startlink{#1}\@@href}%
\providecommand \@@href[1]{\endgroup#1\@@endlink}%
\providecommand \@sanitize@url [0]{\catcode `\\12\catcode `\$12\catcode
  `\&12\catcode `\#12\catcode `\^12\catcode `\_12\catcode `\%12\relax}%
\providecommand \@@startlink[1]{}%
\providecommand \@@endlink[0]{}%
\providecommand \url  [0]{\begingroup\@sanitize@url \@url }%
\providecommand \@url [1]{\endgroup\@href {#1}{\urlprefix }}%
\providecommand \urlprefix  [0]{URL }%
\providecommand \Eprint [0]{\href }%
\providecommand \doibase [0]{http://dx.doi.org/}%
\providecommand \selectlanguage [0]{\@gobble}%
\providecommand \bibinfo  [0]{\@secondoftwo}%
\providecommand \bibfield  [0]{\@secondoftwo}%
\providecommand \translation [1]{[#1]}%
\providecommand \BibitemOpen [0]{}%
\providecommand \bibitemStop [0]{}%
\providecommand \bibitemNoStop [0]{.\EOS\space}%
\providecommand \EOS [0]{\spacefactor3000\relax}%
\providecommand \BibitemShut  [1]{\csname bibitem#1\endcsname}%
\let\auto@bib@innerbib\@empty
%</preamble>
\bibitem [{\citenamefont {Liu}\ \emph {et~al.}(2021)\citenamefont {Liu},
  \citenamefont {Duan}, \citenamefont {Shin}, \citenamefont {Park},
  \citenamefont {Huang},\ and\ \citenamefont {Duan}}]{Liu2021}%
  \BibitemOpen
  \bibfield  {author} {\bibinfo {author} {\bibfnamefont {Y.}~\bibnamefont
  {Liu}}, \bibinfo {author} {\bibfnamefont {X.}~\bibnamefont {Duan}}, \bibinfo
  {author} {\bibfnamefont {H.-J.}\ \bibnamefont {Shin}}, \bibinfo {author}
  {\bibfnamefont {S.}~\bibnamefont {Park}}, \bibinfo {author} {\bibfnamefont
  {Y.}~\bibnamefont {Huang}}, \ and\ \bibinfo {author} {\bibfnamefont
  {X.}~\bibnamefont {Duan}},\ }\href {\doibase 10.1038/s41586-021-03339-z}
  {\bibfield  {journal} {\bibinfo  {journal} {Nature}\ }\textbf {\bibinfo
  {volume} {591}},\ \bibinfo {pages} {43} (\bibinfo {year} {2021})}\BibitemShut
  {NoStop}%
\bibitem [{\citenamefont {Akinwande}\ \emph {et~al.}(2019)\citenamefont
  {Akinwande}, \citenamefont {Huyghebaert}, \citenamefont {Wang}, \citenamefont
  {Serna}, \citenamefont {Goossens}, \citenamefont {Li}, \citenamefont {Wong},\
  and\ \citenamefont {Koppens}}]{Akinwande2019}%
  \BibitemOpen
  \bibfield  {author} {\bibinfo {author} {\bibfnamefont {D.}~\bibnamefont
  {Akinwande}}, \bibinfo {author} {\bibfnamefont {C.}~\bibnamefont
  {Huyghebaert}}, \bibinfo {author} {\bibfnamefont {C.-H.}\ \bibnamefont
  {Wang}}, \bibinfo {author} {\bibfnamefont {M.~I.}\ \bibnamefont {Serna}},
  \bibinfo {author} {\bibfnamefont {S.}~\bibnamefont {Goossens}}, \bibinfo
  {author} {\bibfnamefont {L.-J.}\ \bibnamefont {Li}}, \bibinfo {author}
  {\bibfnamefont {H.-S.~P.}\ \bibnamefont {Wong}}, \ and\ \bibinfo {author}
  {\bibfnamefont {F.~H.~L.}\ \bibnamefont {Koppens}},\ }\href {\doibase
  10.1038/s41586-019-1573-9} {\bibfield  {journal} {\bibinfo  {journal}
  {Nature}\ }\textbf {\bibinfo {volume} {573}},\ \bibinfo {pages} {507}
  (\bibinfo {year} {2019})}\BibitemShut {NoStop}%
\bibitem [{\citenamefont {Liu}\ \emph {et~al.}(2020)\citenamefont {Liu},
  \citenamefont {Chen}, \citenamefont {Wang}, \citenamefont {Liu},
  \citenamefont {Jiang}, \citenamefont {Zhang}, \citenamefont {Liu},\ and\
  \citenamefont {Zhou}}]{Liu2020}%
  \BibitemOpen
  \bibfield  {author} {\bibinfo {author} {\bibfnamefont {C.}~\bibnamefont
  {Liu}}, \bibinfo {author} {\bibfnamefont {H.}~\bibnamefont {Chen}}, \bibinfo
  {author} {\bibfnamefont {S.}~\bibnamefont {Wang}}, \bibinfo {author}
  {\bibfnamefont {Q.}~\bibnamefont {Liu}}, \bibinfo {author} {\bibfnamefont
  {Y.-G.}\ \bibnamefont {Jiang}}, \bibinfo {author} {\bibfnamefont {D.~W.}\
  \bibnamefont {Zhang}}, \bibinfo {author} {\bibfnamefont {M.}~\bibnamefont
  {Liu}}, \ and\ \bibinfo {author} {\bibfnamefont {P.}~\bibnamefont {Zhou}},\
  }\href {\doibase 10.1038/s41565-020-0724-3} {\bibfield  {journal} {\bibinfo
  {journal} {Nature Nanotechnology}\ }\textbf {\bibinfo {volume} {15}},\
  \bibinfo {pages} {545} (\bibinfo {year} {2020})}\BibitemShut {NoStop}%
\bibitem [{\citenamefont {Zhang}\ \emph {et~al.}(2015)\citenamefont {Zhang},
  \citenamefont {Li},\ and\ \citenamefont {Yang}}]{C5NR03895B}%
  \BibitemOpen
  \bibfield  {author} {\bibinfo {author} {\bibfnamefont {R.}~\bibnamefont
  {Zhang}}, \bibinfo {author} {\bibfnamefont {B.}~\bibnamefont {Li}}, \ and\
  \bibinfo {author} {\bibfnamefont {J.}~\bibnamefont {Yang}},\ }\href {\doibase
  10.1039/C5NR03895B} {\bibfield  {journal} {\bibinfo  {journal} {Nanoscale}\
  }\textbf {\bibinfo {volume} {7}},\ \bibinfo {pages} {14062} (\bibinfo {year}
  {2015})}\BibitemShut {NoStop}%
\bibitem [{\citenamefont {Santos}\ \emph {et~al.}(2016)\citenamefont {Santos},
  \citenamefont {Mota}, \citenamefont {Rivelino}, \citenamefont
  {Kakanakova-Georgieva},\ and\ \citenamefont {Gueorguiev}}]{santos}%
  \BibitemOpen
  \bibfield  {author} {\bibinfo {author} {\bibfnamefont {R.}~\bibnamefont
  {Santos}}, \bibinfo {author} {\bibfnamefont {F.}~\bibnamefont {Mota}},
  \bibinfo {author} {\bibfnamefont {R.}~\bibnamefont {Rivelino}}, \bibinfo
  {author} {\bibfnamefont {A.}~\bibnamefont {Kakanakova-Georgieva}}, \ and\
  \bibinfo {author} {\bibfnamefont {G.}~\bibnamefont {Gueorguiev}},\ }\href
  {\doibase 10.1088/0957-4484/27/14/145601} {\bibfield  {journal} {\bibinfo
  {journal} {Nanotechnology}\ }\textbf {\bibinfo {volume} {27}},\ \bibinfo
  {pages} {145601} (\bibinfo {year} {2016})}\BibitemShut {NoStop}%
\bibitem [{\citenamefont {Cantele}\ and\ \citenamefont
  {Ninno}(2017)}]{PhysRevMaterials.1.014002}%
  \BibitemOpen
  \bibfield  {author} {\bibinfo {author} {\bibfnamefont {G.}~\bibnamefont
  {Cantele}}\ and\ \bibinfo {author} {\bibfnamefont {D.}~\bibnamefont
  {Ninno}},\ }\href {\doibase 10.1103/PhysRevMaterials.1.014002} {\bibfield
  {journal} {\bibinfo  {journal} {Phys. Rev. Materials}\ }\textbf {\bibinfo
  {volume} {1}},\ \bibinfo {pages} {014002} (\bibinfo {year}
  {2017})}\BibitemShut {NoStop}%
\bibitem [{\citenamefont {Cao}\ \emph {et~al.}(2018)\citenamefont {Cao},
  \citenamefont {Fatemi}, \citenamefont {Demir}, \citenamefont {Fang},
  \citenamefont {Tomarken}, \citenamefont {Luo}, \citenamefont
  {Sanchez-Yamagishi}, \citenamefont {Watanabe}, \citenamefont {Taniguchi},
  \citenamefont {Kaxiras}, \citenamefont {Ashoori},\ and\ \citenamefont
  {Jarillo-Herrero}}]{Cao2018}%
  \BibitemOpen
  \bibfield  {author} {\bibinfo {author} {\bibfnamefont {Y.}~\bibnamefont
  {Cao}}, \bibinfo {author} {\bibfnamefont {V.}~\bibnamefont {Fatemi}},
  \bibinfo {author} {\bibfnamefont {A.}~\bibnamefont {Demir}}, \bibinfo
  {author} {\bibfnamefont {S.}~\bibnamefont {Fang}}, \bibinfo {author}
  {\bibfnamefont {S.~L.}\ \bibnamefont {Tomarken}}, \bibinfo {author}
  {\bibfnamefont {J.~Y.}\ \bibnamefont {Luo}}, \bibinfo {author} {\bibfnamefont
  {J.~D.}\ \bibnamefont {Sanchez-Yamagishi}}, \bibinfo {author} {\bibfnamefont
  {K.}~\bibnamefont {Watanabe}}, \bibinfo {author} {\bibfnamefont
  {T.}~\bibnamefont {Taniguchi}}, \bibinfo {author} {\bibfnamefont
  {E.}~\bibnamefont {Kaxiras}}, \bibinfo {author} {\bibfnamefont {R.~C.}\
  \bibnamefont {Ashoori}}, \ and\ \bibinfo {author} {\bibfnamefont
  {P.}~\bibnamefont {Jarillo-Herrero}},\ }\href {\doibase 10.1038/nature26154}
  {\bibfield  {journal} {\bibinfo  {journal} {Nature}\ }\textbf {\bibinfo
  {volume} {556}},\ \bibinfo {pages} {80} (\bibinfo {year} {2018})}\BibitemShut
  {NoStop}%
\bibitem [{\citenamefont {Conte}\ \emph {et~al.}(2019)\citenamefont {Conte},
  \citenamefont {Ninno},\ and\ \citenamefont {Cantele}}]{A1}%
  \BibitemOpen
  \bibfield  {author} {\bibinfo {author} {\bibfnamefont {F.}~\bibnamefont
  {Conte}}, \bibinfo {author} {\bibfnamefont {D.}~\bibnamefont {Ninno}}, \ and\
  \bibinfo {author} {\bibfnamefont {G.}~\bibnamefont {Cantele}},\ }\href
  {\doibase 10.1103/PhysRevB.99.155429} {\bibfield  {journal} {\bibinfo
  {journal} {Phys. Rev. B}\ }\textbf {\bibinfo {volume} {99}},\ \bibinfo
  {pages} {155429} (\bibinfo {year} {2019})}\BibitemShut {NoStop}%
\bibitem [{\citenamefont {Lucignano}\ \emph {et~al.}(2019)\citenamefont
  {Lucignano}, \citenamefont {Alf\`e}, \citenamefont {Cataudella},
  \citenamefont {Ninno},\ and\ \citenamefont {Cantele}}]{195419}%
  \BibitemOpen
  \bibfield  {author} {\bibinfo {author} {\bibfnamefont {P.}~\bibnamefont
  {Lucignano}}, \bibinfo {author} {\bibfnamefont {D.}~\bibnamefont {Alf\`e}},
  \bibinfo {author} {\bibfnamefont {V.}~\bibnamefont {Cataudella}}, \bibinfo
  {author} {\bibfnamefont {D.}~\bibnamefont {Ninno}}, \ and\ \bibinfo {author}
  {\bibfnamefont {G.}~\bibnamefont {Cantele}},\ }\href {\doibase
  10.1103/PhysRevB.99.195419} {\bibfield  {journal} {\bibinfo  {journal} {Phys.
  Rev. B}\ }\textbf {\bibinfo {volume} {99}},\ \bibinfo {pages} {195419}
  (\bibinfo {year} {2019})}\BibitemShut {NoStop}%
\bibitem [{\citenamefont {Cantele}\ \emph {et~al.}(2020)\citenamefont
  {Cantele}, \citenamefont {Alf\`e}, \citenamefont {Conte}, \citenamefont
  {Cataudella}, \citenamefont {Ninno},\ and\ \citenamefont
  {Lucignano}}]{TBG_PRR}%
  \BibitemOpen
  \bibfield  {author} {\bibinfo {author} {\bibfnamefont {G.}~\bibnamefont
  {Cantele}}, \bibinfo {author} {\bibfnamefont {D.}~\bibnamefont {Alf\`e}},
  \bibinfo {author} {\bibfnamefont {F.}~\bibnamefont {Conte}}, \bibinfo
  {author} {\bibfnamefont {V.}~\bibnamefont {Cataudella}}, \bibinfo {author}
  {\bibfnamefont {D.}~\bibnamefont {Ninno}}, \ and\ \bibinfo {author}
  {\bibfnamefont {P.}~\bibnamefont {Lucignano}},\ }\href {\doibase
  10.1103/PhysRevResearch.2.043127} {\bibfield  {journal} {\bibinfo  {journal}
  {Phys. Rev. Research}\ }\textbf {\bibinfo {volume} {2}},\ \bibinfo {pages}
  {043127} (\bibinfo {year} {2020})}\BibitemShut {NoStop}%
\bibitem [{\citenamefont {Lebedev}\ \emph {et~al.}(2016)\citenamefont
  {Lebedev}, \citenamefont {Lebedeva}, \citenamefont {Knizhnik},\ and\
  \citenamefont {Popov}}]{C5RA20882C}%
  \BibitemOpen
  \bibfield  {author} {\bibinfo {author} {\bibfnamefont {A.~V.}\ \bibnamefont
  {Lebedev}}, \bibinfo {author} {\bibfnamefont {I.~V.}\ \bibnamefont
  {Lebedeva}}, \bibinfo {author} {\bibfnamefont {A.~A.}\ \bibnamefont
  {Knizhnik}}, \ and\ \bibinfo {author} {\bibfnamefont {A.~M.}\ \bibnamefont
  {Popov}},\ }\href {\doibase 10.1039/C5RA20882C} {\bibfield  {journal}
  {\bibinfo  {journal} {RSC Adv.}\ }\textbf {\bibinfo {volume} {6}},\ \bibinfo
  {pages} {6423} (\bibinfo {year} {2016})}\BibitemShut {NoStop}%
\bibitem [{\citenamefont {Hu}\ \emph {et~al.}(2016)\citenamefont {Hu},
  \citenamefont {Kong}, \citenamefont {Qiao}, \citenamefont {Normand},\ and\
  \citenamefont {Ji}}]{C5NR06293D}%
  \BibitemOpen
  \bibfield  {author} {\bibinfo {author} {\bibfnamefont {Z.-X.}\ \bibnamefont
  {Hu}}, \bibinfo {author} {\bibfnamefont {X.}~\bibnamefont {Kong}}, \bibinfo
  {author} {\bibfnamefont {J.}~\bibnamefont {Qiao}}, \bibinfo {author}
  {\bibfnamefont {B.}~\bibnamefont {Normand}}, \ and\ \bibinfo {author}
  {\bibfnamefont {W.}~\bibnamefont {Ji}},\ }\href {\doibase 10.1039/C5NR06293D}
  {\bibfield  {journal} {\bibinfo  {journal} {Nanoscale}\ }\textbf {\bibinfo
  {volume} {8}},\ \bibinfo {pages} {2740} (\bibinfo {year} {2016})}\BibitemShut
  {NoStop}%
\bibitem [{\citenamefont {Cao}\ \emph {et~al.}(2020)\citenamefont {Cao},
  \citenamefont {Feng}, \citenamefont {Han}, \citenamefont {Gao}, \citenamefont
  {Hue~Ly}, \citenamefont {Xu},\ and\ \citenamefont {Lu}}]{Cao2020}%
  \BibitemOpen
  \bibfield  {author} {\bibinfo {author} {\bibfnamefont {K.}~\bibnamefont
  {Cao}}, \bibinfo {author} {\bibfnamefont {S.}~\bibnamefont {Feng}}, \bibinfo
  {author} {\bibfnamefont {Y.}~\bibnamefont {Han}}, \bibinfo {author}
  {\bibfnamefont {L.}~\bibnamefont {Gao}}, \bibinfo {author} {\bibfnamefont
  {T.}~\bibnamefont {Hue~Ly}}, \bibinfo {author} {\bibfnamefont
  {Z.}~\bibnamefont {Xu}}, \ and\ \bibinfo {author} {\bibfnamefont
  {Y.}~\bibnamefont {Lu}},\ }\href {\doibase 10.1038/s41467-019-14130-0}
  {\bibfield  {journal} {\bibinfo  {journal} {Nature Communications}\ }\textbf
  {\bibinfo {volume} {11}},\ \bibinfo {pages} {284} (\bibinfo {year}
  {2020})}\BibitemShut {NoStop}%
\bibitem [{\citenamefont {Sando}\ \emph {et~al.}(2013)\citenamefont {Sando},
  \citenamefont {Agbelele}, \citenamefont {Rahmedov}, \citenamefont {Liu},
  \citenamefont {Rovillain}, \citenamefont {Toulouse}, \citenamefont {Infante},
  \citenamefont {Pyatakov}, \citenamefont {Fusil}, \citenamefont {Jacquet},
  \citenamefont {Carr{\'e}t{\'e}ro}, \citenamefont {Deranlot}, \citenamefont
  {Lisenkov}, \citenamefont {Wang}, \citenamefont {Le~Breton}, \citenamefont
  {Cazayous}, \citenamefont {Sacuto}, \citenamefont {Juraszek}, \citenamefont
  {Zvezdin}, \citenamefont {Bellaiche}, \citenamefont {Dkhil}, \citenamefont
  {Barth{\'e}l{\'e}my},\ and\ \citenamefont {Bibes}}]{Sando2013}%
  \BibitemOpen
  \bibfield  {author} {\bibinfo {author} {\bibfnamefont {D.}~\bibnamefont
  {Sando}}, \bibinfo {author} {\bibfnamefont {A.}~\bibnamefont {Agbelele}},
  \bibinfo {author} {\bibfnamefont {D.}~\bibnamefont {Rahmedov}}, \bibinfo
  {author} {\bibfnamefont {J.}~\bibnamefont {Liu}}, \bibinfo {author}
  {\bibfnamefont {P.}~\bibnamefont {Rovillain}}, \bibinfo {author}
  {\bibfnamefont {C.}~\bibnamefont {Toulouse}}, \bibinfo {author}
  {\bibfnamefont {I.~C.}\ \bibnamefont {Infante}}, \bibinfo {author}
  {\bibfnamefont {A.~P.}\ \bibnamefont {Pyatakov}}, \bibinfo {author}
  {\bibfnamefont {S.}~\bibnamefont {Fusil}}, \bibinfo {author} {\bibfnamefont
  {E.}~\bibnamefont {Jacquet}}, \bibinfo {author} {\bibfnamefont
  {C.}~\bibnamefont {Carr{\'e}t{\'e}ro}}, \bibinfo {author} {\bibfnamefont
  {C.}~\bibnamefont {Deranlot}}, \bibinfo {author} {\bibfnamefont
  {S.}~\bibnamefont {Lisenkov}}, \bibinfo {author} {\bibfnamefont
  {D.}~\bibnamefont {Wang}}, \bibinfo {author} {\bibfnamefont {J.-M.}\
  \bibnamefont {Le~Breton}}, \bibinfo {author} {\bibfnamefont {M.}~\bibnamefont
  {Cazayous}}, \bibinfo {author} {\bibfnamefont {A.}~\bibnamefont {Sacuto}},
  \bibinfo {author} {\bibfnamefont {J.}~\bibnamefont {Juraszek}}, \bibinfo
  {author} {\bibfnamefont {A.~K.}\ \bibnamefont {Zvezdin}}, \bibinfo {author}
  {\bibfnamefont {L.}~\bibnamefont {Bellaiche}}, \bibinfo {author}
  {\bibfnamefont {B.}~\bibnamefont {Dkhil}}, \bibinfo {author} {\bibfnamefont
  {A.}~\bibnamefont {Barth{\'e}l{\'e}my}}, \ and\ \bibinfo {author}
  {\bibfnamefont {M.}~\bibnamefont {Bibes}},\ }\href {\doibase
  10.1038/nmat3629} {\bibfield  {journal} {\bibinfo  {journal} {Nature
  Materials}\ }\textbf {\bibinfo {volume} {12}},\ \bibinfo {pages} {641}
  (\bibinfo {year} {2013})}\BibitemShut {NoStop}%
\bibitem [{\citenamefont {Fei}\ and\ \citenamefont {Yang}(2014)}]{Fei2014}%
  \BibitemOpen
  \bibfield  {author} {\bibinfo {author} {\bibfnamefont {R.}~\bibnamefont
  {Fei}}\ and\ \bibinfo {author} {\bibfnamefont {L.}~\bibnamefont {Yang}},\
  }\href {\doibase 10.1021/nl500935z} {\bibfield  {journal} {\bibinfo
  {journal} {Nano Letters}\ }\textbf {\bibinfo {volume} {14}},\ \bibinfo
  {pages} {2884} (\bibinfo {year} {2014})}\BibitemShut {NoStop}%
\bibitem [{\citenamefont {Castellanos-Gomez}\ \emph {et~al.}(2013)\citenamefont
  {Castellanos-Gomez}, \citenamefont {Rold{\'a}n}, \citenamefont {Cappelluti},
  \citenamefont {Buscema}, \citenamefont {Guinea}, \citenamefont {van~der
  Zant},\ and\ \citenamefont {Steele}}]{Castellanos-Gomez2013}%
  \BibitemOpen
  \bibfield  {author} {\bibinfo {author} {\bibfnamefont {A.}~\bibnamefont
  {Castellanos-Gomez}}, \bibinfo {author} {\bibfnamefont {R.}~\bibnamefont
  {Rold{\'a}n}}, \bibinfo {author} {\bibfnamefont {E.}~\bibnamefont
  {Cappelluti}}, \bibinfo {author} {\bibfnamefont {M.}~\bibnamefont {Buscema}},
  \bibinfo {author} {\bibfnamefont {F.}~\bibnamefont {Guinea}}, \bibinfo
  {author} {\bibfnamefont {H.~S.~J.}\ \bibnamefont {van~der Zant}}, \ and\
  \bibinfo {author} {\bibfnamefont {G.~A.}\ \bibnamefont {Steele}},\ }\href
  {\doibase 10.1021/nl402875m} {\bibfield  {journal} {\bibinfo  {journal} {Nano
  Letters}\ }\textbf {\bibinfo {volume} {13}},\ \bibinfo {pages} {5361}
  (\bibinfo {year} {2013})}\BibitemShut {NoStop}%
\bibitem [{\citenamefont {Lee}\ \emph {et~al.}(2008)\citenamefont {Lee},
  \citenamefont {Wei}, \citenamefont {Kysar},\ and\ \citenamefont
  {Hone}}]{Lee385}%
  \BibitemOpen
  \bibfield  {author} {\bibinfo {author} {\bibfnamefont {C.}~\bibnamefont
  {Lee}}, \bibinfo {author} {\bibfnamefont {X.}~\bibnamefont {Wei}}, \bibinfo
  {author} {\bibfnamefont {J.~W.}\ \bibnamefont {Kysar}}, \ and\ \bibinfo
  {author} {\bibfnamefont {J.}~\bibnamefont {Hone}},\ }\href {\doibase
  10.1126/science.1157996} {\bibfield  {journal} {\bibinfo  {journal}
  {Science}\ }\textbf {\bibinfo {volume} {321}},\ \bibinfo {pages} {385}
  (\bibinfo {year} {2008})}\BibitemShut {NoStop}%
\bibitem [{\citenamefont {Bertolazzi}\ \emph {et~al.}(2011)\citenamefont
  {Bertolazzi}, \citenamefont {Brivio},\ and\ \citenamefont
  {Kis}}]{Bertolazzi2011}%
  \BibitemOpen
  \bibfield  {author} {\bibinfo {author} {\bibfnamefont {S.}~\bibnamefont
  {Bertolazzi}}, \bibinfo {author} {\bibfnamefont {J.}~\bibnamefont {Brivio}},
  \ and\ \bibinfo {author} {\bibfnamefont {A.}~\bibnamefont {Kis}},\ }\href
  {\doibase 10.1021/nn203879f} {\bibfield  {journal} {\bibinfo  {journal} {ACS
  Nano}\ }\textbf {\bibinfo {volume} {5}},\ \bibinfo {pages} {9703} (\bibinfo
  {year} {2011})}\BibitemShut {NoStop}%
\bibitem [{\citenamefont {Memarzadeh}\ \emph {et~al.}(2021)\citenamefont
  {Memarzadeh}, \citenamefont {Roknabadi}, \citenamefont {Modarresi},
  \citenamefont {Mogulkoc},\ and\ \citenamefont {Rudenko}}]{Memarzadeh_2021}%
  \BibitemOpen
  \bibfield  {author} {\bibinfo {author} {\bibfnamefont {S.}~\bibnamefont
  {Memarzadeh}}, \bibinfo {author} {\bibfnamefont {M.~R.}\ \bibnamefont
  {Roknabadi}}, \bibinfo {author} {\bibfnamefont {M.}~\bibnamefont
  {Modarresi}}, \bibinfo {author} {\bibfnamefont {A.}~\bibnamefont {Mogulkoc}},
  \ and\ \bibinfo {author} {\bibfnamefont {A.~N.}\ \bibnamefont {Rudenko}},\
  }\href {\doibase 10.1088/2053-1583/abf626} {\bibfield  {journal} {\bibinfo
  {journal} {2D Materials}\ }\textbf {\bibinfo {volume} {8}},\ \bibinfo {pages}
  {035022} (\bibinfo {year} {2021})}\BibitemShut {NoStop}%
\bibitem [{\citenamefont {Jiang}\ \emph {et~al.}(2018)\citenamefont {Jiang},
  \citenamefont {Li}, \citenamefont {Wang}, \citenamefont {Mak},\ and\
  \citenamefont {Shan}}]{Jiang2018}%
  \BibitemOpen
  \bibfield  {author} {\bibinfo {author} {\bibfnamefont {S.}~\bibnamefont
  {Jiang}}, \bibinfo {author} {\bibfnamefont {L.}~\bibnamefont {Li}}, \bibinfo
  {author} {\bibfnamefont {Z.}~\bibnamefont {Wang}}, \bibinfo {author}
  {\bibfnamefont {K.~F.}\ \bibnamefont {Mak}}, \ and\ \bibinfo {author}
  {\bibfnamefont {J.}~\bibnamefont {Shan}},\ }\href {\doibase
  10.1038/s41565-018-0135-x} {\bibfield  {journal} {\bibinfo  {journal} {Nature
  Nanotechnology}\ }\textbf {\bibinfo {volume} {13}},\ \bibinfo {pages} {549}
  (\bibinfo {year} {2018})}\BibitemShut {NoStop}%
\bibitem [{\citenamefont {Huang}\ \emph {et~al.}(2018)\citenamefont {Huang},
  \citenamefont {Clark}, \citenamefont {Klein}, \citenamefont {MacNeill},
  \citenamefont {Navarro-Moratalla}, \citenamefont {Seyler}, \citenamefont
  {Wilson}, \citenamefont {McGuire}, \citenamefont {Cobden}, \citenamefont
  {Xiao}, \citenamefont {Yao}, \citenamefont {Jarillo-Herrero},\ and\
  \citenamefont {Xu}}]{Huang2018}%
  \BibitemOpen
  \bibfield  {author} {\bibinfo {author} {\bibfnamefont {B.}~\bibnamefont
  {Huang}}, \bibinfo {author} {\bibfnamefont {G.}~\bibnamefont {Clark}},
  \bibinfo {author} {\bibfnamefont {D.~R.}\ \bibnamefont {Klein}}, \bibinfo
  {author} {\bibfnamefont {D.}~\bibnamefont {MacNeill}}, \bibinfo {author}
  {\bibfnamefont {E.}~\bibnamefont {Navarro-Moratalla}}, \bibinfo {author}
  {\bibfnamefont {K.~L.}\ \bibnamefont {Seyler}}, \bibinfo {author}
  {\bibfnamefont {N.}~\bibnamefont {Wilson}}, \bibinfo {author} {\bibfnamefont
  {M.~A.}\ \bibnamefont {McGuire}}, \bibinfo {author} {\bibfnamefont {D.~H.}\
  \bibnamefont {Cobden}}, \bibinfo {author} {\bibfnamefont {D.}~\bibnamefont
  {Xiao}}, \bibinfo {author} {\bibfnamefont {W.}~\bibnamefont {Yao}}, \bibinfo
  {author} {\bibfnamefont {P.}~\bibnamefont {Jarillo-Herrero}}, \ and\ \bibinfo
  {author} {\bibfnamefont {X.}~\bibnamefont {Xu}},\ }\href {\doibase
  10.1038/s41565-018-0121-3} {\bibfield  {journal} {\bibinfo  {journal} {Nature
  Nanotechnology}\ }\textbf {\bibinfo {volume} {13}},\ \bibinfo {pages} {544}
  (\bibinfo {year} {2018})}\BibitemShut {NoStop}%
\bibitem [{\citenamefont {Li}\ \emph {et~al.}(2019)\citenamefont {Li},
  \citenamefont {Jiang}, \citenamefont {Sivadas}, \citenamefont {Wang},
  \citenamefont {Xu}, \citenamefont {Weber}, \citenamefont {Goldberger},
  \citenamefont {Watanabe}, \citenamefont {Taniguchi}, \citenamefont {Fennie},
  \citenamefont {Fai~Mak},\ and\ \citenamefont {Shan}}]{Li2019}%
  \BibitemOpen
  \bibfield  {author} {\bibinfo {author} {\bibfnamefont {T.}~\bibnamefont
  {Li}}, \bibinfo {author} {\bibfnamefont {S.}~\bibnamefont {Jiang}}, \bibinfo
  {author} {\bibfnamefont {N.}~\bibnamefont {Sivadas}}, \bibinfo {author}
  {\bibfnamefont {Z.}~\bibnamefont {Wang}}, \bibinfo {author} {\bibfnamefont
  {Y.}~\bibnamefont {Xu}}, \bibinfo {author} {\bibfnamefont {D.}~\bibnamefont
  {Weber}}, \bibinfo {author} {\bibfnamefont {J.~E.}\ \bibnamefont
  {Goldberger}}, \bibinfo {author} {\bibfnamefont {K.}~\bibnamefont
  {Watanabe}}, \bibinfo {author} {\bibfnamefont {T.}~\bibnamefont {Taniguchi}},
  \bibinfo {author} {\bibfnamefont {C.~J.}\ \bibnamefont {Fennie}}, \bibinfo
  {author} {\bibfnamefont {K.}~\bibnamefont {Fai~Mak}}, \ and\ \bibinfo
  {author} {\bibfnamefont {J.}~\bibnamefont {Shan}},\ }\href {\doibase
  10.1038/s41563-019-0506-1} {\bibfield  {journal} {\bibinfo  {journal} {Nature
  Materials}\ }\textbf {\bibinfo {volume} {18}},\ \bibinfo {pages} {1303}
  (\bibinfo {year} {2019})}\BibitemShut {NoStop}%
\bibitem [{\citenamefont {Song}\ \emph {et~al.}(2019)\citenamefont {Song},
  \citenamefont {Fei}, \citenamefont {Yankowitz}, \citenamefont {Lin},
  \citenamefont {Jiang}, \citenamefont {Hwangbo}, \citenamefont {Zhang},
  \citenamefont {Sun}, \citenamefont {Taniguchi}, \citenamefont {Watanabe},
  \citenamefont {McGuire}, \citenamefont {Graf}, \citenamefont {Cao},
  \citenamefont {Chu}, \citenamefont {Cobden}, \citenamefont {Dean},
  \citenamefont {Xiao},\ and\ \citenamefont {Xu}}]{Song2019}%
  \BibitemOpen
  \bibfield  {author} {\bibinfo {author} {\bibfnamefont {T.}~\bibnamefont
  {Song}}, \bibinfo {author} {\bibfnamefont {Z.}~\bibnamefont {Fei}}, \bibinfo
  {author} {\bibfnamefont {M.}~\bibnamefont {Yankowitz}}, \bibinfo {author}
  {\bibfnamefont {Z.}~\bibnamefont {Lin}}, \bibinfo {author} {\bibfnamefont
  {Q.}~\bibnamefont {Jiang}}, \bibinfo {author} {\bibfnamefont
  {K.}~\bibnamefont {Hwangbo}}, \bibinfo {author} {\bibfnamefont
  {Q.}~\bibnamefont {Zhang}}, \bibinfo {author} {\bibfnamefont
  {B.}~\bibnamefont {Sun}}, \bibinfo {author} {\bibfnamefont {T.}~\bibnamefont
  {Taniguchi}}, \bibinfo {author} {\bibfnamefont {K.}~\bibnamefont {Watanabe}},
  \bibinfo {author} {\bibfnamefont {M.~A.}\ \bibnamefont {McGuire}}, \bibinfo
  {author} {\bibfnamefont {D.}~\bibnamefont {Graf}}, \bibinfo {author}
  {\bibfnamefont {T.}~\bibnamefont {Cao}}, \bibinfo {author} {\bibfnamefont
  {J.-H.}\ \bibnamefont {Chu}}, \bibinfo {author} {\bibfnamefont {D.~H.}\
  \bibnamefont {Cobden}}, \bibinfo {author} {\bibfnamefont {C.~R.}\
  \bibnamefont {Dean}}, \bibinfo {author} {\bibfnamefont {D.}~\bibnamefont
  {Xiao}}, \ and\ \bibinfo {author} {\bibfnamefont {X.}~\bibnamefont {Xu}},\
  }\href {\doibase 10.1038/s41563-019-0505-2} {\bibfield  {journal} {\bibinfo
  {journal} {Nature Materials}\ }\textbf {\bibinfo {volume} {18}},\ \bibinfo
  {pages} {1298} (\bibinfo {year} {2019})}\BibitemShut {NoStop}%
\bibitem [{\citenamefont {Wu}\ \emph {et~al.}(2019)\citenamefont {Wu},
  \citenamefont {Yu},\ and\ \citenamefont {Yuan}}]{C8CP07067A}%
  \BibitemOpen
  \bibfield  {author} {\bibinfo {author} {\bibfnamefont {Z.}~\bibnamefont
  {Wu}}, \bibinfo {author} {\bibfnamefont {J.}~\bibnamefont {Yu}}, \ and\
  \bibinfo {author} {\bibfnamefont {S.}~\bibnamefont {Yuan}},\ }\href {\doibase
  10.1039/C8CP07067A} {\bibfield  {journal} {\bibinfo  {journal} {Phys. Chem.
  Chem. Phys.}\ }\textbf {\bibinfo {volume} {21}},\ \bibinfo {pages} {7750}
  (\bibinfo {year} {2019})}\BibitemShut {NoStop}%
\bibitem [{\citenamefont {Chittari}\ \emph {et~al.}(2020)\citenamefont
  {Chittari}, \citenamefont {Lee}, \citenamefont {Banerjee}, \citenamefont
  {MacDonald}, \citenamefont {Hwang},\ and\ \citenamefont
  {Jung}}]{PhysRevB.101.085415}%
  \BibitemOpen
  \bibfield  {author} {\bibinfo {author} {\bibfnamefont {B.~L.}\ \bibnamefont
  {Chittari}}, \bibinfo {author} {\bibfnamefont {D.}~\bibnamefont {Lee}},
  \bibinfo {author} {\bibfnamefont {N.}~\bibnamefont {Banerjee}}, \bibinfo
  {author} {\bibfnamefont {A.~H.}\ \bibnamefont {MacDonald}}, \bibinfo {author}
  {\bibfnamefont {E.}~\bibnamefont {Hwang}}, \ and\ \bibinfo {author}
  {\bibfnamefont {J.}~\bibnamefont {Jung}},\ }\href {\doibase
  10.1103/PhysRevB.101.085415} {\bibfield  {journal} {\bibinfo  {journal}
  {Phys. Rev. B}\ }\textbf {\bibinfo {volume} {101}},\ \bibinfo {pages}
  {085415} (\bibinfo {year} {2020})}\BibitemShut {NoStop}%
\bibitem [{\citenamefont {Tai}\ \emph {et~al.}(2020)\citenamefont {Tai},
  \citenamefont {Wu}, \citenamefont {Feng}, \citenamefont {Jiao}, \citenamefont
  {Zhao}, \citenamefont {Lu}, \citenamefont {Sheng},\ and\ \citenamefont
  {Yang}}]{PhysRevB.102.224422}%
  \BibitemOpen
  \bibfield  {author} {\bibinfo {author} {\bibfnamefont {B.}~\bibnamefont
  {Tai}}, \bibinfo {author} {\bibfnamefont {W.}~\bibnamefont {Wu}}, \bibinfo
  {author} {\bibfnamefont {X.}~\bibnamefont {Feng}}, \bibinfo {author}
  {\bibfnamefont {Y.}~\bibnamefont {Jiao}}, \bibinfo {author} {\bibfnamefont
  {J.}~\bibnamefont {Zhao}}, \bibinfo {author} {\bibfnamefont {Y.}~\bibnamefont
  {Lu}}, \bibinfo {author} {\bibfnamefont {X.-L.}\ \bibnamefont {Sheng}}, \
  and\ \bibinfo {author} {\bibfnamefont {S.~A.}\ \bibnamefont {Yang}},\ }\href
  {\doibase 10.1103/PhysRevB.102.224422} {\bibfield  {journal} {\bibinfo
  {journal} {Phys. Rev. B}\ }\textbf {\bibinfo {volume} {102}},\ \bibinfo
  {pages} {224422} (\bibinfo {year} {2020})}\BibitemShut {NoStop}%
\bibitem [{\citenamefont {Dong}\ \emph {et~al.}(2019)\citenamefont {Dong},
  \citenamefont {You}, \citenamefont {Gu},\ and\ \citenamefont
  {Su}}]{PhysRevApplied.12.014020}%
  \BibitemOpen
  \bibfield  {author} {\bibinfo {author} {\bibfnamefont {X.-J.}\ \bibnamefont
  {Dong}}, \bibinfo {author} {\bibfnamefont {J.-Y.}\ \bibnamefont {You}},
  \bibinfo {author} {\bibfnamefont {B.}~\bibnamefont {Gu}}, \ and\ \bibinfo
  {author} {\bibfnamefont {G.}~\bibnamefont {Su}},\ }\href {\doibase
  10.1103/PhysRevApplied.12.014020} {\bibfield  {journal} {\bibinfo  {journal}
  {Phys. Rev. Applied}\ }\textbf {\bibinfo {volume} {12}},\ \bibinfo {pages}
  {014020} (\bibinfo {year} {2019})}\BibitemShut {NoStop}%
\bibitem [{\citenamefont {Huang}\ \emph {et~al.}(2017)\citenamefont {Huang},
  \citenamefont {Clark}, \citenamefont {Navarro-Moratalla}, \citenamefont
  {Klein}, \citenamefont {Cheng}, \citenamefont {Seyler}, \citenamefont
  {Zhong}, \citenamefont {Schmidgall}, \citenamefont {McGuire}, \citenamefont
  {Cobden}, \citenamefont {Yao}, \citenamefont {Xiao}, \citenamefont
  {Jarillo-Herrero},\ and\ \citenamefont {Xu}}]{Huang2017}%
  \BibitemOpen
  \bibfield  {author} {\bibinfo {author} {\bibfnamefont {B.}~\bibnamefont
  {Huang}}, \bibinfo {author} {\bibfnamefont {G.}~\bibnamefont {Clark}},
  \bibinfo {author} {\bibfnamefont {E.}~\bibnamefont {Navarro-Moratalla}},
  \bibinfo {author} {\bibfnamefont {D.~R.}\ \bibnamefont {Klein}}, \bibinfo
  {author} {\bibfnamefont {R.}~\bibnamefont {Cheng}}, \bibinfo {author}
  {\bibfnamefont {K.~L.}\ \bibnamefont {Seyler}}, \bibinfo {author}
  {\bibfnamefont {D.}~\bibnamefont {Zhong}}, \bibinfo {author} {\bibfnamefont
  {E.}~\bibnamefont {Schmidgall}}, \bibinfo {author} {\bibfnamefont {M.~A.}\
  \bibnamefont {McGuire}}, \bibinfo {author} {\bibfnamefont {D.~H.}\
  \bibnamefont {Cobden}}, \bibinfo {author} {\bibfnamefont {W.}~\bibnamefont
  {Yao}}, \bibinfo {author} {\bibfnamefont {D.}~\bibnamefont {Xiao}}, \bibinfo
  {author} {\bibfnamefont {P.}~\bibnamefont {Jarillo-Herrero}}, \ and\ \bibinfo
  {author} {\bibfnamefont {X.}~\bibnamefont {Xu}},\ }\href {\doibase
  10.1038/nature22391} {\bibfield  {journal} {\bibinfo  {journal} {Nature}\
  }\textbf {\bibinfo {volume} {546}},\ \bibinfo {pages} {270} (\bibinfo {year}
  {2017})}\BibitemShut {NoStop}%
\bibitem [{\citenamefont {Blei}\ \emph {et~al.}(2021)\citenamefont {Blei},
  \citenamefont {Lado}, \citenamefont {Song}, \citenamefont {Dey},
  \citenamefont {Erten}, \citenamefont {Pardo}, \citenamefont {Comin},
  \citenamefont {Tongay},\ and\ \citenamefont
  {Botana}}]{doi:10.1063/5.0025658}%
  \BibitemOpen
  \bibfield  {author} {\bibinfo {author} {\bibfnamefont {M.}~\bibnamefont
  {Blei}}, \bibinfo {author} {\bibfnamefont {J.~L.}\ \bibnamefont {Lado}},
  \bibinfo {author} {\bibfnamefont {Q.}~\bibnamefont {Song}}, \bibinfo {author}
  {\bibfnamefont {D.}~\bibnamefont {Dey}}, \bibinfo {author} {\bibfnamefont
  {O.}~\bibnamefont {Erten}}, \bibinfo {author} {\bibfnamefont
  {V.}~\bibnamefont {Pardo}}, \bibinfo {author} {\bibfnamefont
  {R.}~\bibnamefont {Comin}}, \bibinfo {author} {\bibfnamefont
  {S.}~\bibnamefont {Tongay}}, \ and\ \bibinfo {author} {\bibfnamefont {A.~S.}\
  \bibnamefont {Botana}},\ }\href {\doibase 10.1063/5.0025658} {\bibfield
  {journal} {\bibinfo  {journal} {Applied Physics Reviews}\ }\textbf {\bibinfo
  {volume} {8}},\ \bibinfo {pages} {021301} (\bibinfo {year}
  {2021})}\BibitemShut {NoStop}%
\bibitem [{\citenamefont {Gong}\ \emph {et~al.}(2017)\citenamefont {Gong},
  \citenamefont {Li}, \citenamefont {Li}, \citenamefont {Ji}, \citenamefont
  {Stern}, \citenamefont {Xia}, \citenamefont {Cao}, \citenamefont {Bao},
  \citenamefont {Wang}, \citenamefont {Wang}, \citenamefont {Qiu},
  \citenamefont {Cava}, \citenamefont {Louie}, \citenamefont {Xia},\ and\
  \citenamefont {Zhang}}]{Gong2017}%
  \BibitemOpen
  \bibfield  {author} {\bibinfo {author} {\bibfnamefont {C.}~\bibnamefont
  {Gong}}, \bibinfo {author} {\bibfnamefont {L.}~\bibnamefont {Li}}, \bibinfo
  {author} {\bibfnamefont {Z.}~\bibnamefont {Li}}, \bibinfo {author}
  {\bibfnamefont {H.}~\bibnamefont {Ji}}, \bibinfo {author} {\bibfnamefont
  {A.}~\bibnamefont {Stern}}, \bibinfo {author} {\bibfnamefont
  {Y.}~\bibnamefont {Xia}}, \bibinfo {author} {\bibfnamefont {T.}~\bibnamefont
  {Cao}}, \bibinfo {author} {\bibfnamefont {W.}~\bibnamefont {Bao}}, \bibinfo
  {author} {\bibfnamefont {C.}~\bibnamefont {Wang}}, \bibinfo {author}
  {\bibfnamefont {Y.}~\bibnamefont {Wang}}, \bibinfo {author} {\bibfnamefont
  {Z.~Q.}\ \bibnamefont {Qiu}}, \bibinfo {author} {\bibfnamefont {R.~J.}\
  \bibnamefont {Cava}}, \bibinfo {author} {\bibfnamefont {S.~G.}\ \bibnamefont
  {Louie}}, \bibinfo {author} {\bibfnamefont {J.}~\bibnamefont {Xia}}, \ and\
  \bibinfo {author} {\bibfnamefont {X.}~\bibnamefont {Zhang}},\ }\href
  {\doibase 10.1038/nature22060} {\bibfield  {journal} {\bibinfo  {journal}
  {Nature}\ }\textbf {\bibinfo {volume} {546}},\ \bibinfo {pages} {265}
  (\bibinfo {year} {2017})}\BibitemShut {NoStop}%
\bibitem [{\citenamefont {Xiao}\ \emph {et~al.}(2019)\citenamefont {Xiao},
  \citenamefont {Wang}, \citenamefont {Wang}, \citenamefont {Xu}, \citenamefont
  {Liang},\ and\ \citenamefont {Yang}}]{C9CP00850K}%
  \BibitemOpen
  \bibfield  {author} {\bibinfo {author} {\bibfnamefont {H.}~\bibnamefont
  {Xiao}}, \bibinfo {author} {\bibfnamefont {X.}~\bibnamefont {Wang}}, \bibinfo
  {author} {\bibfnamefont {R.}~\bibnamefont {Wang}}, \bibinfo {author}
  {\bibfnamefont {L.}~\bibnamefont {Xu}}, \bibinfo {author} {\bibfnamefont
  {S.}~\bibnamefont {Liang}}, \ and\ \bibinfo {author} {\bibfnamefont
  {C.}~\bibnamefont {Yang}},\ }\href {\doibase 10.1039/C9CP00850K} {\bibfield
  {journal} {\bibinfo  {journal} {Phys. Chem. Chem. Phys.}\ }\textbf {\bibinfo
  {volume} {21}},\ \bibinfo {pages} {11731} (\bibinfo {year}
  {2019})}\BibitemShut {NoStop}%
\bibitem [{\citenamefont {Kan}\ \emph {et~al.}(2014)\citenamefont {Kan},
  \citenamefont {Adhikari},\ and\ \citenamefont {Sun}}]{C3CP55146F}%
  \BibitemOpen
  \bibfield  {author} {\bibinfo {author} {\bibfnamefont {M.}~\bibnamefont
  {Kan}}, \bibinfo {author} {\bibfnamefont {S.}~\bibnamefont {Adhikari}}, \
  and\ \bibinfo {author} {\bibfnamefont {Q.}~\bibnamefont {Sun}},\ }\href
  {\doibase 10.1039/C3CP55146F} {\bibfield  {journal} {\bibinfo  {journal}
  {Phys. Chem. Chem. Phys.}\ }\textbf {\bibinfo {volume} {16}},\ \bibinfo
  {pages} {4990} (\bibinfo {year} {2014})}\BibitemShut {NoStop}%
\bibitem [{\citenamefont {Fuh}\ \emph {et~al.}(2016)\citenamefont {Fuh},
  \citenamefont {Chang}, \citenamefont {Wang}, \citenamefont {Evans},
  \citenamefont {Chantrell},\ and\ \citenamefont {Jeng}}]{Fuh2016}%
  \BibitemOpen
  \bibfield  {author} {\bibinfo {author} {\bibfnamefont {H.-R.}\ \bibnamefont
  {Fuh}}, \bibinfo {author} {\bibfnamefont {C.-R.}\ \bibnamefont {Chang}},
  \bibinfo {author} {\bibfnamefont {Y.-K.}\ \bibnamefont {Wang}}, \bibinfo
  {author} {\bibfnamefont {R.~F.~L.}\ \bibnamefont {Evans}}, \bibinfo {author}
  {\bibfnamefont {R.~W.}\ \bibnamefont {Chantrell}}, \ and\ \bibinfo {author}
  {\bibfnamefont {H.-T.}\ \bibnamefont {Jeng}},\ }\href {\doibase
  10.1038/srep32625} {\bibfield  {journal} {\bibinfo  {journal} {Scientific
  Reports}\ }\textbf {\bibinfo {volume} {6}},\ \bibinfo {pages} {32625}
  (\bibinfo {year} {2016})}\BibitemShut {NoStop}%
\bibitem [{\citenamefont {Conte}\ \emph {et~al.}(2020)\citenamefont {Conte},
  \citenamefont {Ninno},\ and\ \citenamefont {Cantele}}]{Nb3I8_PRR}%
  \BibitemOpen
  \bibfield  {author} {\bibinfo {author} {\bibfnamefont {F.}~\bibnamefont
  {Conte}}, \bibinfo {author} {\bibfnamefont {D.}~\bibnamefont {Ninno}}, \ and\
  \bibinfo {author} {\bibfnamefont {G.}~\bibnamefont {Cantele}},\ }\href
  {\doibase 10.1103/PhysRevResearch.2.033001} {\bibfield  {journal} {\bibinfo
  {journal} {Phys. Rev. Research}\ }\textbf {\bibinfo {volume} {2}},\ \bibinfo
  {pages} {033001} (\bibinfo {year} {2020})}\BibitemShut {NoStop}%
\bibitem [{\citenamefont {\ifmmode \check{Z}\else
  \v{Z}\fi{}uti\ifmmode~\acute{c}\else \'{c}\fi{}}\ \emph
  {et~al.}(2004)\citenamefont {\ifmmode \check{Z}\else
  \v{Z}\fi{}uti\ifmmode~\acute{c}\else \'{c}\fi{}}, \citenamefont {Fabian},\
  and\ \citenamefont {Das~Sarma}}]{RevModPhys.76.323}%
  \BibitemOpen
  \bibfield  {author} {\bibinfo {author} {\bibfnamefont {I.}~\bibnamefont
  {\ifmmode \check{Z}\else \v{Z}\fi{}uti\ifmmode~\acute{c}\else \'{c}\fi{}}},
  \bibinfo {author} {\bibfnamefont {J.}~\bibnamefont {Fabian}}, \ and\ \bibinfo
  {author} {\bibfnamefont {S.}~\bibnamefont {Das~Sarma}},\ }\href {\doibase
  10.1103/RevModPhys.76.323} {\bibfield  {journal} {\bibinfo  {journal} {Rev.
  Mod. Phys.}\ }\textbf {\bibinfo {volume} {76}},\ \bibinfo {pages} {323}
  (\bibinfo {year} {2004})}\BibitemShut {NoStop}%
\bibitem [{\citenamefont {Wolf}\ \emph {et~al.}(2001)\citenamefont {Wolf},
  \citenamefont {Awschalom}, \citenamefont {Buhrman}, \citenamefont {Daughton},
  \citenamefont {von Moln{\'a}r}, \citenamefont {Roukes}, \citenamefont
  {Chtchelkanova},\ and\ \citenamefont {Treger}}]{Wolf1488}%
  \BibitemOpen
  \bibfield  {author} {\bibinfo {author} {\bibfnamefont {S.~A.}\ \bibnamefont
  {Wolf}}, \bibinfo {author} {\bibfnamefont {D.~D.}\ \bibnamefont {Awschalom}},
  \bibinfo {author} {\bibfnamefont {R.~A.}\ \bibnamefont {Buhrman}}, \bibinfo
  {author} {\bibfnamefont {J.~M.}\ \bibnamefont {Daughton}}, \bibinfo {author}
  {\bibfnamefont {S.}~\bibnamefont {von Moln{\'a}r}}, \bibinfo {author}
  {\bibfnamefont {M.~L.}\ \bibnamefont {Roukes}}, \bibinfo {author}
  {\bibfnamefont {A.~Y.}\ \bibnamefont {Chtchelkanova}}, \ and\ \bibinfo
  {author} {\bibfnamefont {D.~M.}\ \bibnamefont {Treger}},\ }\href {\doibase
  10.1126/science.1065389} {\bibfield  {journal} {\bibinfo  {journal}
  {Science}\ }\textbf {\bibinfo {volume} {294}},\ \bibinfo {pages} {1488}
  (\bibinfo {year} {2001})}\BibitemShut {NoStop}%
\bibitem [{\citenamefont {Oh}\ \emph {et~al.}(2020)\citenamefont {Oh},
  \citenamefont {Choi}, \citenamefont {Chae}, \citenamefont {Kim},
  \citenamefont {Jeong}, \citenamefont {Lee}, \citenamefont {Jeon},
  \citenamefont {Kim}, \citenamefont {Nanda}, \citenamefont {Shi},
  \citenamefont {Yi}, \citenamefont {Lee}, \citenamefont {Yu},\ and\
  \citenamefont {Choi}}]{Nb3I8_synthesis}%
  \BibitemOpen
  \bibfield  {author} {\bibinfo {author} {\bibfnamefont {S.}~\bibnamefont
  {Oh}}, \bibinfo {author} {\bibfnamefont {K.~H.}\ \bibnamefont {Choi}},
  \bibinfo {author} {\bibfnamefont {S.}~\bibnamefont {Chae}}, \bibinfo {author}
  {\bibfnamefont {B.~J.}\ \bibnamefont {Kim}}, \bibinfo {author} {\bibfnamefont
  {B.~J.}\ \bibnamefont {Jeong}}, \bibinfo {author} {\bibfnamefont {S.~H.}\
  \bibnamefont {Lee}}, \bibinfo {author} {\bibfnamefont {J.}~\bibnamefont
  {Jeon}}, \bibinfo {author} {\bibfnamefont {Y.}~\bibnamefont {Kim}}, \bibinfo
  {author} {\bibfnamefont {S.~S.}\ \bibnamefont {Nanda}}, \bibinfo {author}
  {\bibfnamefont {L.}~\bibnamefont {Shi}}, \bibinfo {author} {\bibfnamefont
  {D.~K.}\ \bibnamefont {Yi}}, \bibinfo {author} {\bibfnamefont {J.-H.}\
  \bibnamefont {Lee}}, \bibinfo {author} {\bibfnamefont {H.~K.}\ \bibnamefont
  {Yu}}, \ and\ \bibinfo {author} {\bibfnamefont {J.-Y.}\ \bibnamefont
  {Choi}},\ }\href {\doibase https://doi.org/10.1016/j.jallcom.2020.154877}
  {\bibfield  {journal} {\bibinfo  {journal} {Journal of Alloys and Compounds}\
  }\textbf {\bibinfo {volume} {831}},\ \bibinfo {pages} {154877} (\bibinfo
  {year} {2020})}\BibitemShut {NoStop}%
\bibitem [{\citenamefont {Peng}\ \emph {et~al.}(2020)\citenamefont {Peng},
  \citenamefont {Ma}, \citenamefont {Xu}, \citenamefont {He}, \citenamefont
  {Huang},\ and\ \citenamefont {Dai}}]{Nb3I8_valley}%
  \BibitemOpen
  \bibfield  {author} {\bibinfo {author} {\bibfnamefont {R.}~\bibnamefont
  {Peng}}, \bibinfo {author} {\bibfnamefont {Y.}~\bibnamefont {Ma}}, \bibinfo
  {author} {\bibfnamefont {X.}~\bibnamefont {Xu}}, \bibinfo {author}
  {\bibfnamefont {Z.}~\bibnamefont {He}}, \bibinfo {author} {\bibfnamefont
  {B.}~\bibnamefont {Huang}}, \ and\ \bibinfo {author} {\bibfnamefont
  {Y.}~\bibnamefont {Dai}},\ }\href {\doibase 10.1103/PhysRevB.102.035412}
  {\bibfield  {journal} {\bibinfo  {journal} {Phys. Rev. B}\ }\textbf {\bibinfo
  {volume} {102}},\ \bibinfo {pages} {035412} (\bibinfo {year}
  {2020})}\BibitemShut {NoStop}%
\bibitem [{\citenamefont {Regmi}\ \emph {et~al.}(2022)\citenamefont {Regmi},
  \citenamefont {Fernando}, \citenamefont {Zhao}, \citenamefont {Sakhya},
  \citenamefont {Dhakal}, \citenamefont {Elius}, \citenamefont {Vazquez},
  \citenamefont {Denlinger}, \citenamefont {Yang}, \citenamefont {Chu},
  \citenamefont {Xu}, \citenamefont {Cao},\ and\ \citenamefont
  {Neupane}}]{arxiv.2203.10547}%
  \BibitemOpen
  \bibfield  {author} {\bibinfo {author} {\bibfnamefont {S.}~\bibnamefont
  {Regmi}}, \bibinfo {author} {\bibfnamefont {T.~W.}\ \bibnamefont {Fernando}},
  \bibinfo {author} {\bibfnamefont {Y.}~\bibnamefont {Zhao}}, \bibinfo {author}
  {\bibfnamefont {A.~P.}\ \bibnamefont {Sakhya}}, \bibinfo {author}
  {\bibfnamefont {G.}~\bibnamefont {Dhakal}}, \bibinfo {author} {\bibfnamefont
  {I.~B.}\ \bibnamefont {Elius}}, \bibinfo {author} {\bibfnamefont
  {H.}~\bibnamefont {Vazquez}}, \bibinfo {author} {\bibfnamefont {J.~D.}\
  \bibnamefont {Denlinger}}, \bibinfo {author} {\bibfnamefont {J.}~\bibnamefont
  {Yang}}, \bibinfo {author} {\bibfnamefont {J.-H.}\ \bibnamefont {Chu}},
  \bibinfo {author} {\bibfnamefont {X.}~\bibnamefont {Xu}}, \bibinfo {author}
  {\bibfnamefont {T.}~\bibnamefont {Cao}}, \ and\ \bibinfo {author}
  {\bibfnamefont {M.}~\bibnamefont {Neupane}},\ }\href@noop {} {\bibfield
  {journal} {\bibinfo  {journal} {arXiv}\ ,\ \bibinfo {pages}
  {arxiv.2203.10547}} (\bibinfo {year} {2022})}\BibitemShut {NoStop}%
\bibitem [{\citenamefont {Sun}\ \emph {et~al.}(2022)\citenamefont {Sun},
  \citenamefont {Zhou}, \citenamefont {Wang}, \citenamefont {Kumar},
  \citenamefont {Geng}, \citenamefont {Yue}, \citenamefont {Han}, \citenamefont
  {Haraguchi}, \citenamefont {Shimada}, \citenamefont {Cheng}, \citenamefont
  {Chen}, \citenamefont {Shi}, \citenamefont {Wu}, \citenamefont {Meng},\ and\
  \citenamefont {Feng}}]{Sun2022}%
  \BibitemOpen
  \bibfield  {author} {\bibinfo {author} {\bibfnamefont {Z.}~\bibnamefont
  {Sun}}, \bibinfo {author} {\bibfnamefont {H.}~\bibnamefont {Zhou}}, \bibinfo
  {author} {\bibfnamefont {C.}~\bibnamefont {Wang}}, \bibinfo {author}
  {\bibfnamefont {S.}~\bibnamefont {Kumar}}, \bibinfo {author} {\bibfnamefont
  {D.}~\bibnamefont {Geng}}, \bibinfo {author} {\bibfnamefont {S.}~\bibnamefont
  {Yue}}, \bibinfo {author} {\bibfnamefont {X.}~\bibnamefont {Han}}, \bibinfo
  {author} {\bibfnamefont {Y.}~\bibnamefont {Haraguchi}}, \bibinfo {author}
  {\bibfnamefont {K.}~\bibnamefont {Shimada}}, \bibinfo {author} {\bibfnamefont
  {P.}~\bibnamefont {Cheng}}, \bibinfo {author} {\bibfnamefont
  {L.}~\bibnamefont {Chen}}, \bibinfo {author} {\bibfnamefont {Y.}~\bibnamefont
  {Shi}}, \bibinfo {author} {\bibfnamefont {K.}~\bibnamefont {Wu}}, \bibinfo
  {author} {\bibfnamefont {S.}~\bibnamefont {Meng}}, \ and\ \bibinfo {author}
  {\bibfnamefont {B.}~\bibnamefont {Feng}},\ }\href {\doibase
  10.1021/acs.nanolett.2c00778} {\bibfield  {journal} {\bibinfo  {journal}
  {Nano Letters}\ } (\bibinfo {year} {2022}),\
  10.1021/acs.nanolett.2c00778}\BibitemShut {NoStop}%
\bibitem [{\citenamefont {Giannozzi}\ \emph {et~al.}(2009)\citenamefont
  {Giannozzi}, \citenamefont {Baroni}, \citenamefont {Bonini}, \citenamefont
  {Calandra}, \citenamefont {Car}, \citenamefont {Cavazzoni}, \citenamefont
  {Ceresoli}, \citenamefont {Chiarotti}, \citenamefont {Cococcioni},
  \citenamefont {Dabo}, \citenamefont {Corso}, \citenamefont {de~Gironcoli},
  \citenamefont {Fabris}, \citenamefont {Fratesi}, \citenamefont {Gebauer},
  \citenamefont {Gerstmann}, \citenamefont {Gougoussis}, \citenamefont
  {Kokalj}, \citenamefont {Lazzeri}, \citenamefont {Martin-Samos},
  \citenamefont {Marzari}, \citenamefont {Mauri}, \citenamefont {Mazzarello},
  \citenamefont {Paolini}, \citenamefont {Pasquarello}, \citenamefont
  {Paulatto}, \citenamefont {Sbraccia}, \citenamefont {Scandolo}, \citenamefont
  {Sclauzero}, \citenamefont {Seitsonen}, \citenamefont {Smogunov},
  \citenamefont {Umari},\ and\ \citenamefont {Wentzcovitch}}]{Giannozzi_2009}%
  \BibitemOpen
  \bibfield  {author} {\bibinfo {author} {\bibfnamefont {P.}~\bibnamefont
  {Giannozzi}}, \bibinfo {author} {\bibfnamefont {S.}~\bibnamefont {Baroni}},
  \bibinfo {author} {\bibfnamefont {N.}~\bibnamefont {Bonini}}, \bibinfo
  {author} {\bibfnamefont {M.}~\bibnamefont {Calandra}}, \bibinfo {author}
  {\bibfnamefont {R.}~\bibnamefont {Car}}, \bibinfo {author} {\bibfnamefont
  {C.}~\bibnamefont {Cavazzoni}}, \bibinfo {author} {\bibfnamefont
  {D.}~\bibnamefont {Ceresoli}}, \bibinfo {author} {\bibfnamefont {G.~L.}\
  \bibnamefont {Chiarotti}}, \bibinfo {author} {\bibfnamefont {M.}~\bibnamefont
  {Cococcioni}}, \bibinfo {author} {\bibfnamefont {I.}~\bibnamefont {Dabo}},
  \bibinfo {author} {\bibfnamefont {A.~D.}\ \bibnamefont {Corso}}, \bibinfo
  {author} {\bibfnamefont {S.}~\bibnamefont {de~Gironcoli}}, \bibinfo {author}
  {\bibfnamefont {S.}~\bibnamefont {Fabris}}, \bibinfo {author} {\bibfnamefont
  {G.}~\bibnamefont {Fratesi}}, \bibinfo {author} {\bibfnamefont
  {R.}~\bibnamefont {Gebauer}}, \bibinfo {author} {\bibfnamefont
  {U.}~\bibnamefont {Gerstmann}}, \bibinfo {author} {\bibfnamefont
  {C.}~\bibnamefont {Gougoussis}}, \bibinfo {author} {\bibfnamefont
  {A.}~\bibnamefont {Kokalj}}, \bibinfo {author} {\bibfnamefont
  {M.}~\bibnamefont {Lazzeri}}, \bibinfo {author} {\bibfnamefont
  {L.}~\bibnamefont {Martin-Samos}}, \bibinfo {author} {\bibfnamefont
  {N.}~\bibnamefont {Marzari}}, \bibinfo {author} {\bibfnamefont
  {F.}~\bibnamefont {Mauri}}, \bibinfo {author} {\bibfnamefont
  {R.}~\bibnamefont {Mazzarello}}, \bibinfo {author} {\bibfnamefont
  {S.}~\bibnamefont {Paolini}}, \bibinfo {author} {\bibfnamefont
  {A.}~\bibnamefont {Pasquarello}}, \bibinfo {author} {\bibfnamefont
  {L.}~\bibnamefont {Paulatto}}, \bibinfo {author} {\bibfnamefont
  {C.}~\bibnamefont {Sbraccia}}, \bibinfo {author} {\bibfnamefont
  {S.}~\bibnamefont {Scandolo}}, \bibinfo {author} {\bibfnamefont
  {G.}~\bibnamefont {Sclauzero}}, \bibinfo {author} {\bibfnamefont {A.~P.}\
  \bibnamefont {Seitsonen}}, \bibinfo {author} {\bibfnamefont {A.}~\bibnamefont
  {Smogunov}}, \bibinfo {author} {\bibfnamefont {P.}~\bibnamefont {Umari}}, \
  and\ \bibinfo {author} {\bibfnamefont {R.~M.}\ \bibnamefont {Wentzcovitch}},\
  }\href@noop {} {\bibfield  {journal} {\bibinfo  {journal} {Journal of
  Physics: Condensed Matter}\ }\textbf {\bibinfo {volume} {21}},\ \bibinfo
  {pages} {395502} (\bibinfo {year} {2009})}\BibitemShut {NoStop}%
\bibitem [{\citenamefont {Giannozzi}\ \emph {et~al.}(2017)\citenamefont
  {Giannozzi}, \citenamefont {Andreussi}, \citenamefont {Brumme}, \citenamefont
  {Bunau}, \citenamefont {Nardelli}, \citenamefont {Calandra}, \citenamefont
  {Car}, \citenamefont {Cavazzoni}, \citenamefont {Ceresoli}, \citenamefont
  {Cococcioni}, \citenamefont {Colonna}, \citenamefont {Carnimeo},
  \citenamefont {Corso}, \citenamefont {de~Gironcoli}, \citenamefont {Delugas},
  \citenamefont {DiStasio}, \citenamefont {Ferretti}, \citenamefont {Floris},
  \citenamefont {Fratesi}, \citenamefont {Fugallo}, \citenamefont {Gebauer},
  \citenamefont {Gerstmann}, \citenamefont {Giustino}, \citenamefont {Gorni},
  \citenamefont {Jia}, \citenamefont {Kawamura}, \citenamefont {Ko},
  \citenamefont {Kokalj}, \citenamefont {Kü{\c{c}}ükbenli}, \citenamefont
  {Lazzeri}, \citenamefont {Marsili}, \citenamefont {Marzari}, \citenamefont
  {Mauri}, \citenamefont {Nguyen}, \citenamefont {Nguyen}, \citenamefont {de-la
  Roza}, \citenamefont {Paulatto}, \citenamefont {Ponc{\'{e}}}, \citenamefont
  {Rocca}, \citenamefont {Sabatini}, \citenamefont {Santra}, \citenamefont
  {Schlipf}, \citenamefont {Seitsonen}, \citenamefont {Smogunov}, \citenamefont
  {Timrov}, \citenamefont {Thonhauser}, \citenamefont {Umari}, \citenamefont
  {Vast}, \citenamefont {Wu},\ and\ \citenamefont {Baroni}}]{Giannozzi_2017}%
  \BibitemOpen
  \bibfield  {author} {\bibinfo {author} {\bibfnamefont {P.}~\bibnamefont
  {Giannozzi}}, \bibinfo {author} {\bibfnamefont {O.}~\bibnamefont
  {Andreussi}}, \bibinfo {author} {\bibfnamefont {T.}~\bibnamefont {Brumme}},
  \bibinfo {author} {\bibfnamefont {O.}~\bibnamefont {Bunau}}, \bibinfo
  {author} {\bibfnamefont {M.~B.}\ \bibnamefont {Nardelli}}, \bibinfo {author}
  {\bibfnamefont {M.}~\bibnamefont {Calandra}}, \bibinfo {author}
  {\bibfnamefont {R.}~\bibnamefont {Car}}, \bibinfo {author} {\bibfnamefont
  {C.}~\bibnamefont {Cavazzoni}}, \bibinfo {author} {\bibfnamefont
  {D.}~\bibnamefont {Ceresoli}}, \bibinfo {author} {\bibfnamefont
  {M.}~\bibnamefont {Cococcioni}}, \bibinfo {author} {\bibfnamefont
  {N.}~\bibnamefont {Colonna}}, \bibinfo {author} {\bibfnamefont
  {I.}~\bibnamefont {Carnimeo}}, \bibinfo {author} {\bibfnamefont {A.~D.}\
  \bibnamefont {Corso}}, \bibinfo {author} {\bibfnamefont {S.}~\bibnamefont
  {de~Gironcoli}}, \bibinfo {author} {\bibfnamefont {P.}~\bibnamefont
  {Delugas}}, \bibinfo {author} {\bibfnamefont {R.~A.}\ \bibnamefont
  {DiStasio}}, \bibinfo {author} {\bibfnamefont {A.}~\bibnamefont {Ferretti}},
  \bibinfo {author} {\bibfnamefont {A.}~\bibnamefont {Floris}}, \bibinfo
  {author} {\bibfnamefont {G.}~\bibnamefont {Fratesi}}, \bibinfo {author}
  {\bibfnamefont {G.}~\bibnamefont {Fugallo}}, \bibinfo {author} {\bibfnamefont
  {R.}~\bibnamefont {Gebauer}}, \bibinfo {author} {\bibfnamefont
  {U.}~\bibnamefont {Gerstmann}}, \bibinfo {author} {\bibfnamefont
  {F.}~\bibnamefont {Giustino}}, \bibinfo {author} {\bibfnamefont
  {T.}~\bibnamefont {Gorni}}, \bibinfo {author} {\bibfnamefont
  {J.}~\bibnamefont {Jia}}, \bibinfo {author} {\bibfnamefont {M.}~\bibnamefont
  {Kawamura}}, \bibinfo {author} {\bibfnamefont {H.-Y.}\ \bibnamefont {Ko}},
  \bibinfo {author} {\bibfnamefont {A.}~\bibnamefont {Kokalj}}, \bibinfo
  {author} {\bibfnamefont {E.}~\bibnamefont {Kü{\c{c}}ükbenli}}, \bibinfo
  {author} {\bibfnamefont {M.}~\bibnamefont {Lazzeri}}, \bibinfo {author}
  {\bibfnamefont {M.}~\bibnamefont {Marsili}}, \bibinfo {author} {\bibfnamefont
  {N.}~\bibnamefont {Marzari}}, \bibinfo {author} {\bibfnamefont
  {F.}~\bibnamefont {Mauri}}, \bibinfo {author} {\bibfnamefont {N.~L.}\
  \bibnamefont {Nguyen}}, \bibinfo {author} {\bibfnamefont {H.-V.}\
  \bibnamefont {Nguyen}}, \bibinfo {author} {\bibfnamefont {A.~O.}\
  \bibnamefont {de-la Roza}}, \bibinfo {author} {\bibfnamefont
  {L.}~\bibnamefont {Paulatto}}, \bibinfo {author} {\bibfnamefont
  {S.}~\bibnamefont {Ponc{\'{e}}}}, \bibinfo {author} {\bibfnamefont
  {D.}~\bibnamefont {Rocca}}, \bibinfo {author} {\bibfnamefont
  {R.}~\bibnamefont {Sabatini}}, \bibinfo {author} {\bibfnamefont
  {B.}~\bibnamefont {Santra}}, \bibinfo {author} {\bibfnamefont
  {M.}~\bibnamefont {Schlipf}}, \bibinfo {author} {\bibfnamefont {A.~P.}\
  \bibnamefont {Seitsonen}}, \bibinfo {author} {\bibfnamefont {A.}~\bibnamefont
  {Smogunov}}, \bibinfo {author} {\bibfnamefont {I.}~\bibnamefont {Timrov}},
  \bibinfo {author} {\bibfnamefont {T.}~\bibnamefont {Thonhauser}}, \bibinfo
  {author} {\bibfnamefont {P.}~\bibnamefont {Umari}}, \bibinfo {author}
  {\bibfnamefont {N.}~\bibnamefont {Vast}}, \bibinfo {author} {\bibfnamefont
  {X.}~\bibnamefont {Wu}}, \ and\ \bibinfo {author} {\bibfnamefont
  {S.}~\bibnamefont {Baroni}},\ }\href {\doibase 10.1088/1361-648x/aa8f79}
  {\bibfield  {journal} {\bibinfo  {journal} {Journal of Physics: Condensed
  Matter}\ }\textbf {\bibinfo {volume} {29}},\ \bibinfo {pages} {465901}
  (\bibinfo {year} {2017})}\BibitemShut {NoStop}%
\bibitem [{\citenamefont {Giannozzi}\ \emph {et~al.}(2020)\citenamefont
  {Giannozzi}, \citenamefont {Baseggio}, \citenamefont {Bonfà}, \citenamefont
  {Brunato}, \citenamefont {Car}, \citenamefont {Carnimeo}, \citenamefont
  {Cavazzoni}, \citenamefont {de~Gironcoli}, \citenamefont {Delugas},
  \citenamefont {Ferrari~Ruffino}, \citenamefont {Ferretti}, \citenamefont
  {Marzari}, \citenamefont {Timrov}, \citenamefont {Urru},\ and\ \citenamefont
  {Baroni}}]{Giannozzi_2020}%
  \BibitemOpen
  \bibfield  {author} {\bibinfo {author} {\bibfnamefont {P.}~\bibnamefont
  {Giannozzi}}, \bibinfo {author} {\bibfnamefont {O.}~\bibnamefont {Baseggio}},
  \bibinfo {author} {\bibfnamefont {P.}~\bibnamefont {Bonfà}}, \bibinfo
  {author} {\bibfnamefont {D.}~\bibnamefont {Brunato}}, \bibinfo {author}
  {\bibfnamefont {R.}~\bibnamefont {Car}}, \bibinfo {author} {\bibfnamefont
  {I.}~\bibnamefont {Carnimeo}}, \bibinfo {author} {\bibfnamefont
  {C.}~\bibnamefont {Cavazzoni}}, \bibinfo {author} {\bibfnamefont
  {S.}~\bibnamefont {de~Gironcoli}}, \bibinfo {author} {\bibfnamefont
  {P.}~\bibnamefont {Delugas}}, \bibinfo {author} {\bibfnamefont
  {F.}~\bibnamefont {Ferrari~Ruffino}}, \bibinfo {author} {\bibfnamefont
  {A.}~\bibnamefont {Ferretti}}, \bibinfo {author} {\bibfnamefont
  {N.}~\bibnamefont {Marzari}}, \bibinfo {author} {\bibfnamefont
  {I.}~\bibnamefont {Timrov}}, \bibinfo {author} {\bibfnamefont
  {A.}~\bibnamefont {Urru}}, \ and\ \bibinfo {author} {\bibfnamefont
  {S.}~\bibnamefont {Baroni}},\ }\href {\doibase 10.1063/5.0005082} {\bibfield
  {journal} {\bibinfo  {journal} {The Journal of Chemical Physics}\ }\textbf
  {\bibinfo {volume} {152}},\ \bibinfo {pages} {154105} (\bibinfo {year}
  {2020})}\BibitemShut {NoStop}%
\bibitem [{\citenamefont {Corso}(2014)}]{DALCORSO2014337}%
  \BibitemOpen
  \bibfield  {author} {\bibinfo {author} {\bibfnamefont {A.~D.}\ \bibnamefont
  {Corso}},\ }\href@noop {} {\bibfield  {journal} {\bibinfo  {journal}
  {Computational Materials Science}\ }\textbf {\bibinfo {volume} {95}},\
  \bibinfo {pages} {337 } (\bibinfo {year} {2014})}\BibitemShut {NoStop}%
\bibitem [{\citenamefont {Perdew}\ \emph {et~al.}(1996)\citenamefont {Perdew},
  \citenamefont {Burke},\ and\ \citenamefont
  {Ernzerhof}}]{PhysRevLett.77.3865}%
  \BibitemOpen
  \bibfield  {author} {\bibinfo {author} {\bibfnamefont {J.~P.}\ \bibnamefont
  {Perdew}}, \bibinfo {author} {\bibfnamefont {K.}~\bibnamefont {Burke}}, \
  and\ \bibinfo {author} {\bibfnamefont {M.}~\bibnamefont {Ernzerhof}},\ }\href
  {\doibase 10.1103/PhysRevLett.77.3865} {\bibfield  {journal} {\bibinfo
  {journal} {Phys. Rev. Lett.}\ }\textbf {\bibinfo {volume} {77}},\ \bibinfo
  {pages} {3865} (\bibinfo {year} {1996})}\BibitemShut {NoStop}%
\bibitem [{PSE()}]{PSEUDO}%
  \BibitemOpen
  \href@noop {} {}\bibinfo {note} {We used the scalar relativistic
  pseudopotentials \textsf{Nb.pbe-spn-kjpaw\_psl.1.0.0.UPF} and
  \textsf{I.pbe-n-kjpaw\_psl.1.0.0.UPF} (for Nb and I atoms, respectively) from
  the Quantum ESPRESSO pseudopotential data base:
  http://www.quantum-espresso.org/pseudopotentials}\BibitemShut {NoStop}%
\bibitem [{\citenamefont {Monkhorst}\ and\ \citenamefont
  {Pack}(1976)}]{Monkhorst}%
  \BibitemOpen
  \bibfield  {author} {\bibinfo {author} {\bibfnamefont {H.~J.}\ \bibnamefont
  {Monkhorst}}\ and\ \bibinfo {author} {\bibfnamefont {J.~D.}\ \bibnamefont
  {Pack}},\ }\href {\doibase 10.1103/PhysRevB.13.5188} {\bibfield  {journal}
  {\bibinfo  {journal} {Phys. Rev. B}\ }\textbf {\bibinfo {volume} {13}},\
  \bibinfo {pages} {5188} (\bibinfo {year} {1976})}\BibitemShut {NoStop}%
\bibitem [{\citenamefont {Hamada}(2014)}]{B86R}%
  \BibitemOpen
  \bibfield  {author} {\bibinfo {author} {\bibfnamefont {I.}~\bibnamefont
  {Hamada}},\ }\href {\doibase 10.1103/PhysRevB.89.121103} {\bibfield
  {journal} {\bibinfo  {journal} {Phys. Rev. B}\ }\textbf {\bibinfo {volume}
  {89}},\ \bibinfo {pages} {121103(R)} (\bibinfo {year} {2014})}\BibitemShut
  {NoStop}%
\bibitem [{\citenamefont {Jiang}\ \emph {et~al.}(2017)\citenamefont {Jiang},
  \citenamefont {Liang}, \citenamefont {Meng}, \citenamefont {Yang},
  \citenamefont {Tan}, \citenamefont {Sun},\ and\ \citenamefont
  {Chen}}]{C6NR07231C}%
  \BibitemOpen
  \bibfield  {author} {\bibinfo {author} {\bibfnamefont {J.}~\bibnamefont
  {Jiang}}, \bibinfo {author} {\bibfnamefont {Q.}~\bibnamefont {Liang}},
  \bibinfo {author} {\bibfnamefont {R.}~\bibnamefont {Meng}}, \bibinfo {author}
  {\bibfnamefont {Q.}~\bibnamefont {Yang}}, \bibinfo {author} {\bibfnamefont
  {C.}~\bibnamefont {Tan}}, \bibinfo {author} {\bibfnamefont {X.}~\bibnamefont
  {Sun}}, \ and\ \bibinfo {author} {\bibfnamefont {X.}~\bibnamefont {Chen}},\
  }\href@noop {} {\bibfield  {journal} {\bibinfo  {journal} {Nanoscale}\
  }\textbf {\bibinfo {volume} {9}},\ \bibinfo {pages} {2992} (\bibinfo {year}
  {2017})}\BibitemShut {NoStop}%
\bibitem [{\citenamefont {Anisimov}\ \emph {et~al.}(1991)\citenamefont
  {Anisimov}, \citenamefont {Zaanen},\ and\ \citenamefont {Andersen}}]{LDAU1}%
  \BibitemOpen
  \bibfield  {author} {\bibinfo {author} {\bibfnamefont {V.~I.}\ \bibnamefont
  {Anisimov}}, \bibinfo {author} {\bibfnamefont {J.}~\bibnamefont {Zaanen}}, \
  and\ \bibinfo {author} {\bibfnamefont {O.~K.}\ \bibnamefont {Andersen}},\
  }\href {\doibase 10.1103/PhysRevB.44.943} {\bibfield  {journal} {\bibinfo
  {journal} {Phys. Rev. B}\ }\textbf {\bibinfo {volume} {44}},\ \bibinfo
  {pages} {943} (\bibinfo {year} {1991})}\BibitemShut {NoStop}%
\bibitem [{\citenamefont {Anisimov}\ \emph {et~al.}(1993)\citenamefont
  {Anisimov}, \citenamefont {Solovyev}, \citenamefont {Korotin}, \citenamefont
  {Czyzyk},\ and\ \citenamefont {Sawatzky}}]{LDAU2}%
  \BibitemOpen
  \bibfield  {author} {\bibinfo {author} {\bibfnamefont {V.~I.}\ \bibnamefont
  {Anisimov}}, \bibinfo {author} {\bibfnamefont {I.~V.}\ \bibnamefont
  {Solovyev}}, \bibinfo {author} {\bibfnamefont {M.~A.}\ \bibnamefont
  {Korotin}}, \bibinfo {author} {\bibfnamefont {M.~T.}\ \bibnamefont {Czyzyk}},
  \ and\ \bibinfo {author} {\bibfnamefont {G.~A.}\ \bibnamefont {Sawatzky}},\
  }\href {\doibase 10.1103/PhysRevB.48.16929} {\bibfield  {journal} {\bibinfo
  {journal} {Phys. Rev. B}\ }\textbf {\bibinfo {volume} {48}},\ \bibinfo
  {pages} {16929} (\bibinfo {year} {1993})}\BibitemShut {NoStop}%
\bibitem [{\citenamefont {Anisimov}\ \emph {et~al.}(1997)\citenamefont
  {Anisimov}, \citenamefont {Aryasetiawan},\ and\ \citenamefont
  {Lichtenstein}}]{LDAU3}%
  \BibitemOpen
  \bibfield  {author} {\bibinfo {author} {\bibfnamefont {V.~I.}\ \bibnamefont
  {Anisimov}}, \bibinfo {author} {\bibfnamefont {F.}~\bibnamefont
  {Aryasetiawan}}, \ and\ \bibinfo {author} {\bibfnamefont {A.~I.}\
  \bibnamefont {Lichtenstein}},\ }\href {\doibase 10.1088/0953-8984/9/4/002}
  {\bibfield  {journal} {\bibinfo  {journal} {Journal of Physics: Condensed
  Matter}\ }\textbf {\bibinfo {volume} {9}},\ \bibinfo {pages} {767} (\bibinfo
  {year} {1997})}\BibitemShut {NoStop}%
\bibitem [{\citenamefont {Cococcioni}\ and\ \citenamefont
  {de~Gironcoli}(2005)}]{PhysRevB.71.035105}%
  \BibitemOpen
  \bibfield  {author} {\bibinfo {author} {\bibfnamefont {M.}~\bibnamefont
  {Cococcioni}}\ and\ \bibinfo {author} {\bibfnamefont {S.}~\bibnamefont
  {de~Gironcoli}},\ }\href@noop {} {\bibfield  {journal} {\bibinfo  {journal}
  {Phys. Rev. B}\ }\textbf {\bibinfo {volume} {71}},\ \bibinfo {pages} {035105}
  (\bibinfo {year} {2005})}\BibitemShut {NoStop}%
\bibitem [{\citenamefont {Broyden}(1970)}]{B}%
  \BibitemOpen
  \bibfield  {author} {\bibinfo {author} {\bibfnamefont {C.~G.}\ \bibnamefont
  {Broyden}},\ }\href@noop {} {\bibfield  {journal} {\bibinfo  {journal} {IMA
  Journal of Applied Mathematics}\ }\textbf {\bibinfo {volume} {6}},\ \bibinfo
  {pages} {222} (\bibinfo {year} {1970})}\BibitemShut {NoStop}%
\bibitem [{\citenamefont {Fletcher}(1970)}]{F}%
  \BibitemOpen
  \bibfield  {author} {\bibinfo {author} {\bibfnamefont {R.}~\bibnamefont
  {Fletcher}},\ }\href@noop {} {\bibfield  {journal} {\bibinfo  {journal} {The
  Computer Journal}\ }\textbf {\bibinfo {volume} {13}},\ \bibinfo {pages} {317}
  (\bibinfo {year} {1970})}\BibitemShut {NoStop}%
\bibitem [{\citenamefont {Goldfarb}(1970)}]{G}%
  \BibitemOpen
  \bibfield  {author} {\bibinfo {author} {\bibfnamefont {D.}~\bibnamefont
  {Goldfarb}},\ }\href@noop {} {\bibfield  {journal} {\bibinfo  {journal}
  {Mathematics of Computation}\ }\textbf {\bibinfo {volume} {24}},\ \bibinfo
  {pages} {23} (\bibinfo {year} {1970})}\BibitemShut {NoStop}%
\bibitem [{\citenamefont {Shanno}(1970)}]{S}%
  \BibitemOpen
  \bibfield  {author} {\bibinfo {author} {\bibfnamefont {D.~F.}\ \bibnamefont
  {Shanno}},\ }\href@noop {} {\bibfield  {journal} {\bibinfo  {journal}
  {Mathematics of Computation}\ }\textbf {\bibinfo {volume} {24}},\ \bibinfo
  {pages} {647} (\bibinfo {year} {1970})}\BibitemShut {NoStop}%
\bibitem [{\citenamefont {Hulliger}(1976)}]{hulliger}%
  \BibitemOpen
  \bibfield  {author} {\bibinfo {author} {\bibfnamefont {F.}~\bibnamefont
  {Hulliger}},\ }\href@noop {} {\emph {\bibinfo {title} {Structural Chemistry
  of Layer-Type Phases}}},\ edited by\ \bibinfo {editor} {\bibfnamefont
  {F.}~\bibnamefont {L\'evy}},\ Physics and Chemistry of Materials with A\
  (\bibinfo  {publisher} {Springer Netherlands},\ \bibinfo {year} {1976})\ p.\
  \bibinfo {pages} {392}\BibitemShut {NoStop}%
\bibitem [{\citenamefont {Kim}\ \emph {et~al.}(2019)\citenamefont {Kim},
  \citenamefont {Jeong}, \citenamefont {Oh}, \citenamefont {Chae},
  \citenamefont {Choi}, \citenamefont {Nanda}, \citenamefont {Nasir},
  \citenamefont {Lee}, \citenamefont {Kim}, \citenamefont {Lim}, \citenamefont
  {Chi}, \citenamefont {Choi}, \citenamefont {Hong}, \citenamefont {Yi},
  \citenamefont {Yu}, \citenamefont {Lee},\ and\ \citenamefont
  {Choi}}]{https://doi.org/10.1002/pssr.201800448}%
  \BibitemOpen
  \bibfield  {author} {\bibinfo {author} {\bibfnamefont {B.~J.}\ \bibnamefont
  {Kim}}, \bibinfo {author} {\bibfnamefont {B.~J.}\ \bibnamefont {Jeong}},
  \bibinfo {author} {\bibfnamefont {S.}~\bibnamefont {Oh}}, \bibinfo {author}
  {\bibfnamefont {S.}~\bibnamefont {Chae}}, \bibinfo {author} {\bibfnamefont
  {K.~H.}\ \bibnamefont {Choi}}, \bibinfo {author} {\bibfnamefont {S.~S.}\
  \bibnamefont {Nanda}}, \bibinfo {author} {\bibfnamefont {T.}~\bibnamefont
  {Nasir}}, \bibinfo {author} {\bibfnamefont {S.~H.}\ \bibnamefont {Lee}},
  \bibinfo {author} {\bibfnamefont {K.-W.}\ \bibnamefont {Kim}}, \bibinfo
  {author} {\bibfnamefont {H.~K.}\ \bibnamefont {Lim}}, \bibinfo {author}
  {\bibfnamefont {L.}~\bibnamefont {Chi}}, \bibinfo {author} {\bibfnamefont
  {I.~J.}\ \bibnamefont {Choi}}, \bibinfo {author} {\bibfnamefont {M.-K.}\
  \bibnamefont {Hong}}, \bibinfo {author} {\bibfnamefont {D.~K.}\ \bibnamefont
  {Yi}}, \bibinfo {author} {\bibfnamefont {H.~K.}\ \bibnamefont {Yu}}, \bibinfo
  {author} {\bibfnamefont {J.-H.}\ \bibnamefont {Lee}}, \ and\ \bibinfo
  {author} {\bibfnamefont {J.-Y.}\ \bibnamefont {Choi}},\ }\href@noop {}
  {\bibfield  {journal} {\bibinfo  {journal} {physica status solidi (RRL) –
  Rapid Research Letters}\ }\textbf {\bibinfo {volume} {13}},\ \bibinfo {pages}
  {1800448} (\bibinfo {year} {2019})}\BibitemShut {NoStop}%
\bibitem [{\citenamefont {Xiao}\ \emph {et~al.}(2021)\citenamefont {Xiao},
  \citenamefont {Chen},\ and\ \citenamefont {Tong}}]{PhysRevResearch.3.013027}%
  \BibitemOpen
  \bibfield  {author} {\bibinfo {author} {\bibfnamefont {F.}~\bibnamefont
  {Xiao}}, \bibinfo {author} {\bibfnamefont {K.}~\bibnamefont {Chen}}, \ and\
  \bibinfo {author} {\bibfnamefont {Q.}~\bibnamefont {Tong}},\ }\href@noop {}
  {\bibfield  {journal} {\bibinfo  {journal} {Phys. Rev. Research}\ }\textbf
  {\bibinfo {volume} {3}},\ \bibinfo {pages} {013027} (\bibinfo {year}
  {2021})}\BibitemShut {NoStop}%
\bibitem [{\citenamefont {Sivadas}\ \emph {et~al.}(2018)\citenamefont
  {Sivadas}, \citenamefont {Okamoto}, \citenamefont {Xu}, \citenamefont
  {Fennie},\ and\ \citenamefont {Xiao}}]{doi:10.1021/acs.nanolett.8b03321}%
  \BibitemOpen
  \bibfield  {author} {\bibinfo {author} {\bibfnamefont {N.}~\bibnamefont
  {Sivadas}}, \bibinfo {author} {\bibfnamefont {S.}~\bibnamefont {Okamoto}},
  \bibinfo {author} {\bibfnamefont {X.}~\bibnamefont {Xu}}, \bibinfo {author}
  {\bibfnamefont {C.~J.}\ \bibnamefont {Fennie}}, \ and\ \bibinfo {author}
  {\bibfnamefont {D.}~\bibnamefont {Xiao}},\ }\href@noop {} {\bibfield
  {journal} {\bibinfo  {journal} {Nano Letters}\ }\textbf {\bibinfo {volume}
  {18}},\ \bibinfo {pages} {7658} (\bibinfo {year} {2018})}\BibitemShut
  {NoStop}%
\bibitem [{\citenamefont {Akram}\ \emph {et~al.}(2021)\citenamefont {Akram},
  \citenamefont {LaBollita}, \citenamefont {Dey}, \citenamefont {Kapeghian},
  \citenamefont {Erten},\ and\ \citenamefont
  {Botana}}]{doi:10.1021/acs.nanolett.1c02096}%
  \BibitemOpen
  \bibfield  {author} {\bibinfo {author} {\bibfnamefont {M.}~\bibnamefont
  {Akram}}, \bibinfo {author} {\bibfnamefont {H.}~\bibnamefont {LaBollita}},
  \bibinfo {author} {\bibfnamefont {D.}~\bibnamefont {Dey}}, \bibinfo {author}
  {\bibfnamefont {J.}~\bibnamefont {Kapeghian}}, \bibinfo {author}
  {\bibfnamefont {O.}~\bibnamefont {Erten}}, \ and\ \bibinfo {author}
  {\bibfnamefont {A.~S.}\ \bibnamefont {Botana}},\ }\href@noop {} {\bibfield
  {journal} {\bibinfo  {journal} {Nano Letters}\ }\textbf {\bibinfo {volume}
  {21}},\ \bibinfo {pages} {6633} (\bibinfo {year} {2021})}\BibitemShut
  {NoStop}%
\bibitem [{\citenamefont {Ci}\ \emph {et~al.}(2022)\citenamefont {Ci},
  \citenamefont {Yang}, \citenamefont {Xue}, \citenamefont {Yang},
  \citenamefont {Lv}, \citenamefont {Wang}, \citenamefont {Li},\ and\
  \citenamefont {Xu}}]{Ci2022}%
  \BibitemOpen
  \bibfield  {author} {\bibinfo {author} {\bibfnamefont {W.}~\bibnamefont
  {Ci}}, \bibinfo {author} {\bibfnamefont {H.}~\bibnamefont {Yang}}, \bibinfo
  {author} {\bibfnamefont {W.}~\bibnamefont {Xue}}, \bibinfo {author}
  {\bibfnamefont {R.}~\bibnamefont {Yang}}, \bibinfo {author} {\bibfnamefont
  {B.}~\bibnamefont {Lv}}, \bibinfo {author} {\bibfnamefont {P.}~\bibnamefont
  {Wang}}, \bibinfo {author} {\bibfnamefont {R.-W.}\ \bibnamefont {Li}}, \ and\
  \bibinfo {author} {\bibfnamefont {X.-H.}\ \bibnamefont {Xu}},\ }\href
  {\doibase 10.1007/s12274-022-4400-9} {\bibfield  {journal} {\bibinfo
  {journal} {Nano Research}\ } (\bibinfo {year} {2022}),\
  10.1007/s12274-022-4400-9}\BibitemShut {NoStop}%
\bibitem [{\citenamefont {Cantele}\ \emph {et~al.}(2009)\citenamefont
  {Cantele}, \citenamefont {Lee}, \citenamefont {Ninno},\ and\ \citenamefont
  {Marzari}}]{doi:10.1021/nl901557x}%
  \BibitemOpen
  \bibfield  {author} {\bibinfo {author} {\bibfnamefont {G.}~\bibnamefont
  {Cantele}}, \bibinfo {author} {\bibfnamefont {Y.-S.}\ \bibnamefont {Lee}},
  \bibinfo {author} {\bibfnamefont {D.}~\bibnamefont {Ninno}}, \ and\ \bibinfo
  {author} {\bibfnamefont {N.}~\bibnamefont {Marzari}},\ }\href@noop {}
  {\bibfield  {journal} {\bibinfo  {journal} {Nano Letters}\ }\textbf {\bibinfo
  {volume} {9}},\ \bibinfo {pages} {3425} (\bibinfo {year} {2009})}\BibitemShut
  {NoStop}%
\bibitem [{\citenamefont {Jiang}\ \emph {et~al.}(2021)\citenamefont {Jiang},
  \citenamefont {Liu}, \citenamefont {Xing}, \citenamefont {Liu}, \citenamefont
  {Guo}, \citenamefont {Liu},\ and\ \citenamefont
  {Zhao}}]{doi:10.1063/5.0039979}%
  \BibitemOpen
  \bibfield  {author} {\bibinfo {author} {\bibfnamefont {X.}~\bibnamefont
  {Jiang}}, \bibinfo {author} {\bibfnamefont {Q.}~\bibnamefont {Liu}}, \bibinfo
  {author} {\bibfnamefont {J.}~\bibnamefont {Xing}}, \bibinfo {author}
  {\bibfnamefont {N.}~\bibnamefont {Liu}}, \bibinfo {author} {\bibfnamefont
  {Y.}~\bibnamefont {Guo}}, \bibinfo {author} {\bibfnamefont {Z.}~\bibnamefont
  {Liu}}, \ and\ \bibinfo {author} {\bibfnamefont {J.}~\bibnamefont {Zhao}},\
  }\href@noop {} {\bibfield  {journal} {\bibinfo  {journal} {Applied Physics
  Reviews}\ }\textbf {\bibinfo {volume} {8}},\ \bibinfo {pages} {031305}
  (\bibinfo {year} {2021})}\BibitemShut {NoStop}%
\bibitem [{\citenamefont {Zhao}\ \emph {et~al.}(2021)\citenamefont {Zhao},
  \citenamefont {Li}, \citenamefont {Zeng}, \citenamefont {Huang},
  \citenamefont {Yun}, \citenamefont {Zhang},\ and\ \citenamefont
  {Hou}}]{https://doi.org/10.1002/sstr.202100077}%
  \BibitemOpen
  \bibfield  {author} {\bibinfo {author} {\bibfnamefont {Z.}~\bibnamefont
  {Zhao}}, \bibinfo {author} {\bibfnamefont {W.}~\bibnamefont {Li}}, \bibinfo
  {author} {\bibfnamefont {Y.}~\bibnamefont {Zeng}}, \bibinfo {author}
  {\bibfnamefont {X.}~\bibnamefont {Huang}}, \bibinfo {author} {\bibfnamefont
  {C.}~\bibnamefont {Yun}}, \bibinfo {author} {\bibfnamefont {B.}~\bibnamefont
  {Zhang}}, \ and\ \bibinfo {author} {\bibfnamefont {Y.}~\bibnamefont {Hou}},\
  }\href@noop {} {\bibfield  {journal} {\bibinfo  {journal} {Small Structures}\
  }\textbf {\bibinfo {volume} {2}},\ \bibinfo {pages} {2100077} (\bibinfo
  {year} {2021})}\BibitemShut {NoStop}%
\bibitem [{\citenamefont {Zhang}\ \emph {et~al.}(2017)\citenamefont {Zhang},
  \citenamefont {Huang},\ and\ \citenamefont {Cazalilla}}]{Zhang_2017}%
  \BibitemOpen
  \bibfield  {author} {\bibinfo {author} {\bibfnamefont {X.-P.}\ \bibnamefont
  {Zhang}}, \bibinfo {author} {\bibfnamefont {C.}~\bibnamefont {Huang}}, \ and\
  \bibinfo {author} {\bibfnamefont {M.~A.}\ \bibnamefont {Cazalilla}},\ }\href
  {\doibase 10.1088/2053-1583/aa5e9b} {\bibfield  {journal} {\bibinfo
  {journal} {2D Materials}\ }\textbf {\bibinfo {volume} {4}},\ \bibinfo {pages}
  {024007} (\bibinfo {year} {2017})}\BibitemShut {NoStop}%
\bibitem [{\citenamefont {Yang}\ \emph {et~al.}(2019)\citenamefont {Yang},
  \citenamefont {Li}, \citenamefont {Liu}, \citenamefont {Li},\ and\
  \citenamefont {Mao}}]{C9CP02404B}%
  \BibitemOpen
  \bibfield  {author} {\bibinfo {author} {\bibfnamefont {G.}~\bibnamefont
  {Yang}}, \bibinfo {author} {\bibfnamefont {J.}~\bibnamefont {Li}}, \bibinfo
  {author} {\bibfnamefont {Z.}~\bibnamefont {Liu}}, \bibinfo {author}
  {\bibfnamefont {C.}~\bibnamefont {Li}}, \ and\ \bibinfo {author}
  {\bibfnamefont {X.}~\bibnamefont {Mao}},\ }\href@noop {} {\bibfield
  {journal} {\bibinfo  {journal} {Phys. Chem. Chem. Phys.}\ }\textbf {\bibinfo
  {volume} {21}},\ \bibinfo {pages} {15151} (\bibinfo {year}
  {2019})}\BibitemShut {NoStop}%
\bibitem [{\citenamefont {Guan}\ and\ \citenamefont
  {Ni}(2020)}]{doi:10.1021/acsami.0c13988}%
  \BibitemOpen
  \bibfield  {author} {\bibinfo {author} {\bibfnamefont {Z.}~\bibnamefont
  {Guan}}\ and\ \bibinfo {author} {\bibfnamefont {S.}~\bibnamefont {Ni}},\
  }\href@noop {} {\bibfield  {journal} {\bibinfo  {journal} {ACS Applied
  Materials \& Interfaces}\ }\textbf {\bibinfo {volume} {12}},\ \bibinfo
  {pages} {53067} (\bibinfo {year} {2020})}\BibitemShut {NoStop}%
\end{thebibliography}%

\end{document}